\documentclass[twocolumn]{aastex6}

\usepackage{hyperref}
\usepackage{amsmath}

\usepackage{color}
\usepackage{soul}

\usepackage{graphicx}
\usepackage{epstopdf}

\usepackage{booktabs,caption}
\usepackage[flushleft]{threeparttable}
\makeatletter
\def\env@matrix{\hskip -\arraycolsep 
	\let\@ifnextchar\new@ifnextchar
	\array{*{\c@MaxMatrixCols}c}}
\makeatother
\usepackage{blindtext}

\shorttitle{Photoevaporation  of circumbinary PPDs}
\shortauthors{Shadmehri, Ghoreyshi, and Alipour} 

\begin{document}

\title{Evolution  of circumbinary protoplanetary disks with photoevaporative winds driven by External Far Ultraviolet Radiation}

\author{M. Shadmehri$^{1}$, S. M. Ghoreyshi$^{1,2 \star}$, N. Alipour$^{1, \star}$}
\affil{
$^{1}$Department of Physics, Faculty of Science, Golestan University, Gorgan 49138-15739, Iran\\
$^{2}$ Research Institute for Astronomy and Astrophysics of Maragha (RIAAM), Maragha, P.O. Box: 55134-441, Iran
}
\email{m.shadmehri@gu.ac.ir (MS)}
\email{$^{\star}$Iran Science Elites Federation postdoctoral fellow}

\begin{abstract}
Lifetimes of  protoplanetary disks (PPDs) are believed to be severely constrained by material depleting mechanisms, including photoevaporative winds due to the host star radiation or external radiation sources. Most previous studies focused on exploring the role of the winds in the exposed PPDs with a single star; however, exploring the evolution of the circumbinary disks with the photoevaporative winds driven by the host star radiation and external radiation sources deserves further investigation. In this study, we investigate the evolution of the circumbinary PPDs with the photoevaporative winds induced by external far ultraviolet (FUV) radiation field. We show that this mass-loss process can significantly constrain properties of a circumbinary PPD, including its lifetime, mass and radius. The lifetime of a circumbinary PPD, for instance, is found by a factor of about two longer than a similar circumstellar disk and this  enhancement strongly depends on the viscosity parameter. But our model shows that viscosity dependence of the disk lifetime in the circumbinary case is more pronounced compared to the circumstellar case. We also show that dispersal of a circumbinary PPD occurs over a longer time as the disk temperature distribution becomes steeper. Our results also imply that dead zone in a photoevaporative circumbinary PPD extends over a larger radial range in comparison to a circumstellar disk counterpart.  We also show that our calculations are in agreement with the observed circumbinary PPDs orbiting equal-mass binaries.   
\end{abstract}


\section{Introduction}

 Since planet formation time-scale should not exceed lifetimes of the protoplanetary disks (PPDs), constraining their lifetime plays an essential role in the current theories of planet formation \citep[for recent reviews, e.g.,][]{Armitage11,Ercolano17}. Magnetically driven winds \citep{Bland82} or photoevaporative mechanisms \citep[e.g.,][]{Alex2006,Gorti2009} and mass accretion onto the central star or already formed planets are efficient mass depletion processes that will eventually lead to a PPD dispersal. The relative importance of these mass-loss processes, however, strongly depends upon the PPDs physical properties. While magnetized winds are known to be effective in extraction of mass  angular momentum \citep[][]{Bai2016,wang2017ApJ}, photoevaporation due to the radiation field of the central star \citep[e.g.,][]{Alex2006,Gorti2009} or its ambient stars are efficient in mass removal \citep{Ander13}. the typical lifetime of a PPD is estimated to be  less than 10 Myr \citep{Kraus2012, Ander13, Kimura2016, Li2017}. Furthermore, properties of the molecular cloud cores within which PPDs are thought to be formed can dramatically affect the structure and evolution of the PPDs \citep{Li2016,Xiao18}.
 
The mass-loss rate driven by radiation field of the host star or external sources, as evaporative agents in the disk erosion, is a key quantity in constructing disk models with the photoevaporative winds. Primary focus of most previous studies was to elaborate role of the internal radiation sources, including X-ray \citep[e.g.,][]{Owen10,Owen11} and UV-radiation \citep[e.g.,][]{Alex2006} in  dispersal of the isolated PPDs. But PPDs residing in populated regions that contain OB stars are also exposed to their ambient radiation field \citep{Clarke2007,Ander13}. \cite{Ander13} (hereafter; AAC2013) studied evolution of a viscous disk with the photoevaporation due to far ultraviolet (FUV) radiation flux from external stars using existing photoevaporative models \citep{Adams2004}. In order to explore the relative importance of the internal and externals radiation fields in  disk erosion, they also considered X-ray photoevaporation due to the host star and found that external sources are  dominant in the dispersal of a PPD with a solar-mass host star. A PPD lifetime, its mass and radius, therefore, were constrained severely due to the external FUV radiation field (AAC2013).

Following recent discoveries of the circumbinary planets \citep{Doyle11,Orosz12,Schwamb13}, there is a growing interest to understand the evolution of the circumbinary PPDs. Following  the standard disc model \citep{SS73}, various circumbinary disk models that incorporate  binary torque have been developed in recent years \cite[e.g.,][]{Shapiro2010,Kocsis2012,Martin13,Rafikov2013}.  Numerical models are also used to explore circumbinary disk evolution and the binary-disk interactions \citep[e.g.,][]{Yu2002,MacFadyen2008,Roedig2012,Ragusa2016,Miranda2017,Tang2017}.  Exploring disks around individual components of a binary system is another line of research.  \cite{Rosotti18}, for instance, studied the evolution of the discs around components of a binary system with photoevaporation by X-rays from the respective star.

Although many authors have studied the structure of circumbinary disks by performing numerical simulations, presenting analytical models are still useful due to simplicity in interpreting results. These analytical models constructed based on the standard accretion disk model \citep{SS73} with a parameterization of the binary torque and the associated heating term \citep[e.g.,][]{Shapiro2010,Kocsis2012,Rafikov2013,shadmehri2015,Vartan16}. Note that some of these models intended to be used for studying disks orbiting supermassive binary black holes \citep{Shapiro2010,Kocsis2012,Rafikov2013}. In the context of circumbinary PPDs, just recently, \cite[][hereafter; VGR2016]{Vartan16} developed a  disk model without winds to explore its steady-state structure  and evolution via analytical and numerical solutions. Their analysis showed that binary torque in the innermost region has a profound effect on the entire disk structure. They showed a circumbinary disk evolves with a significantly reduced accretion rate in its inner edge in comparison to a similar disk with a single star. A circumbinary disk, therefore, evolves on a longer time-scale in comparison to a circumstellar  disk counterpart.

In the light of this finding and prominent role of the photoevaporative winds in shortening a disk lifetime, it is worthwhile exploring the structure of a circumbinary disk in the presence of this mass-loss process. This problem has been addressed by \cite{Alexander12} who studied the evolution of a circumbinary disk with photoevaporative winds due to the radiation field of the host star. \cite{Alexander12} found that a photoevaporative circumbinary disk evolves with a larger surface density comparing to a disk counterpart with a single star. \cite{Alexander12} primarily studied role of the {\it internal} radiation field due to the host star in erosion of the circumbinary disks, whereas we plan to investigate constraints on a circumbinary PPD quantities in the presence of winds driven by {\it external} radiation field such as ambient stars which is a dominant evaporative agent in the disks with a solar-mass host star according to AAC2013. However, we also provide a comparative study by including photoevaporation due to the internal and external radiation sources. We perform a detailed parameter study over a broad range of the model parameters and a comparison is made between the obtained results and the observed properties of some circumbinary disks. 

In section 2, we present basic equations which are generalized forms of the standard disk model to include the binary torque and the photoevaporative wind mass-loss. In section 3, we investigate the  evolution of the photoevaporative circumbinary PPDs corresponding to various sets of the model parameters. We then compare the obtained results with some of the observed circumbinary PPDs in section 4. Finally, we discuss the model and summarize our main findings in section 5. 

\section{Basic Equations}

A circumbinary PPD is modeled as a thin disk with a binary system at its center. The orbital plane of the binary with the primary and secondary masses $M_{\rm p}$ and $M_{\rm s}$ and the semimajor axis $a_{\rm b}$ is assumed to be coplanar with the disk. The mass ratio of the binary components is $q\equiv M_{\rm s} /M_{\rm p} \leq 1$. Although the disk is subject to a time-varying gravitational potential due to the binary orbital motion, as an approximation, we assume that the disk is rotating in the potential arising from the total mass, i.e., $M_{\rm c}=M_{\rm p}+M_{\rm s}$. Disk rotation profile, therefore, is Keplerian with the angular velocity $\Omega=(GM_{\rm c}/r^3)^{1/2}$. All disk quantities, furthermore, are assumed to be dependent only on the radial distance $r$ and time $t$.

Under these assumptions and following the standard approach for constructing a thin disk model \citep{SS73}, the surface density evolution equation for a circumbinary disk in the presence of the wind mass-loss is
\begin{equation}
\label{eq:main} 
\frac{\partial \Sigma}{\partial t}=\frac{1}{r}\frac{\partial}{\partial
 r}\left [3r^{1/2}\frac{\partial}{\partial r}\left ( \nu\Sigma r^{1/2} \right )-\frac{2\Lambda \Sigma}{\Omega} \right ]-\dot{\Sigma}_{\rm w},
\end{equation}
where $\nu$ is the turbulent viscosity and $\Lambda$ is the specific angular momentum injection rate by the binary. The rate of the wind mass-loss is denoted by $\dot{\Sigma}_{\rm w}$. When the angular momentum injection rate is set to zero and the wind mass-loss is neglected, the  equation (\ref{eq:main}) above reduces to the surface density evolution equation for a disk surrounding a single star. In the presence of the wind, i.e. $\dot{\Sigma}_{\rm w}\neq 0$, the equation (\ref{eq:main}) describes evolution of a single star disk with wind mass-loss. Our focus, instead, is to explore the evolution of a circumbinary disk subject to the evaporative winds. In doing so, we have to specify three important quantities.

Disk turbulence, however, is thought to be driven by the fluid instabilities, including magnetorotational instability \citep[MRI;][]{Balbus91} or gravitational instability \citep[for a recent review, e.g.,][]{Kratter2016} depending upon the disk properties. While a PPD inner region is subject to MRI as the main source of the turbulence, the outer part of a massive enough PPD is gravitationally unstable. Therefore, the first key quantity is the turbulent viscosity $\nu$ where its functional dependence on the disk quantities is defined in an {\it ad-hoc} fashion within the framework of the standard disk model \citep{SS73}. Although describing turbulence in terms of an effective viscosity is a  primitive approach due to the non-linear and chaotic nature of this complex phenomenon, in the standard thin disk model all these complexities are simplified when the azimuthal-radial component of the stress tensor is assumed to be proportional to the pressure. This approach leads to a commonly used relation for the turbulent viscosity, i.e. $\nu=\alpha c_{\rm s}^2 /\Omega$, where $\alpha<1$  and $c_{\rm s}$ are the viscosity parameter and the sound speed, respectively. The sound speed is written in terms of the disk midplane temperature $T$, i.e. $c_{\rm s}=\sqrt{k_{\rm B} T/\mu}$ where $k_{\rm B}$ is the Boltzmann constant and $\mu=2.1 $ m$_{\rm H}$ is the mean molecular weight and m$_{\rm H}$ is hydrogen mass.

Our adopted viscosity prescription depends on the disk temperature and the radial distance. Thus, we  do need another relation to close set of the model equations. Energy balance equation is adequate for that purpose, however, we can instead prescribe disk temperature as a power-law function of the radial distance, i.e. 
\begin{equation}
T= 300 \left (\frac{r}{1 {\rm AU}} \right )^{-s} {\rm K},
\end{equation}
where the temperature exponent is $0< s \leq 1$. When the energy budget of a disk is dominated by the stellar irradiation instead of the viscous heating, the temperature exponent becomes $s=1/2$ \citep[e.g.,][]{Frank2002} which is our standard value. But we  will explore the role of this exponent on the evolution of the circumbinary disks by considering different values for it (see Figure \ref{fig:f5}).

\begin{table}
     \centering
    \caption{List of important quantities in this article.}\label{tab:1}
     \begin{tabular}{ll}
        \hline \hline
        Parameter  & \( {\rm Symbol} \) \\
        \hline
        \\
                    \multicolumn{2}{c}{Disk Parameters}     \\  
        viscosity parameter & $\alpha$ \\ 
        disk midplane temperature & $T$\\  
        mean molecular weight & $\mu$\\   
        temperature exponent & $s$ \\
        \\
        \hline
        \\
                    \multicolumn{2}{c}{Binary Parameters}     \\
       primary mass & $M_{\rm p}$ \\ 
       secondary mass & $M_{\rm s}$\\ 
       total mass & $M_{\rm c}$\\ 
       mass ratio & $q$ \\ 
       semimajor axis & $a_{\rm b}$\\ 
       binary torque coefficient & $f$\\
       \\
        \hline 
        \\
                    \multicolumn{2}{c}{Wind Parameters}     \\
        flux level & $G_0$\\
        gas number density & $n_d$ \\ 
        critical radius& $r_g$\\
        heated gas temperature & $T_{\rm ph}$ \\
        \\
        \hline            
     \end{tabular}
     \vspace{1ex}
\end{table}

The second key quantity is the rate of angular momentum injection by the binary to the disk which is denoted by $\Lambda$(${\rm {cm}}^2 {\rm s}^{-2}$) in equation (\ref{eq:main}). Although the angular momentum injection is restricted to a narrow annulus in the inner region of a disk and it rapidly decreases with the distance, this process is able to affect global structure of a disk. If the mass ratio $q$ is not very small, the binary torque is able to clear out a cavity in the disk center where its size depends on the binary separation \citep{MacFadyen2008} and eccentricity \citep{Pelupessy2013}. The binary tidal torque is approximated as \citep{Armitage2002}
\begin{equation}\label{eq:torque}
\Lambda(r)={\rm sgn}(r-a_{\rm b})\frac{fq^2GM_{\rm c}}{2r}(\frac{a_{\rm b}}{\Delta_{\rm p}})^{4},
\end{equation}
where $f$ is a dimensionless normalization factor. Furthermore, $\Delta_{\rm p}$ is defined by $\Delta_{\rm p}={\rm max}(H,r- a_{\rm b})$ where $H=c_{\rm s} /\Omega $ is the disk scale height. In the literature, different values for $f$ in a range between 0.001 and 1 have been considered \citep[e.g.,][]{Armitage2002,Alexander12,Martin13}. In the case of circumbinary PPDs, for instance, \cite{Alexander12} and \cite{Martin13} assumed that $f=1$, but VGR2016 adopted a much lower value, i.e., $f=0.001$. We, however, note that the binary torque relation in VGR2016 is different from our equation (\ref{eq:torque}) which has been implemented in most prior works. In modeling disks around supermassive binary black holes, on the other hand, \cite{Armitage2002} proposed that $f=0.01$. We  adopt $f=1$ as our standard value; however, other values within the reported range will be considered to explore role of this parameter in evolution of the photoevaporative circumbinary PPDs (see Figure \ref{fig:f8}). We note that the binary torque equation (\ref{eq:torque}) is not quite appropriate for the binaries with nearly equal-mass components. Using this relation in our model, however, is justified by its simplicity and large uncertainties in other model parameters \citep[see also section 3 in][]{Vartan16}.

\begin{figure*}
\includegraphics[scale=1.0]{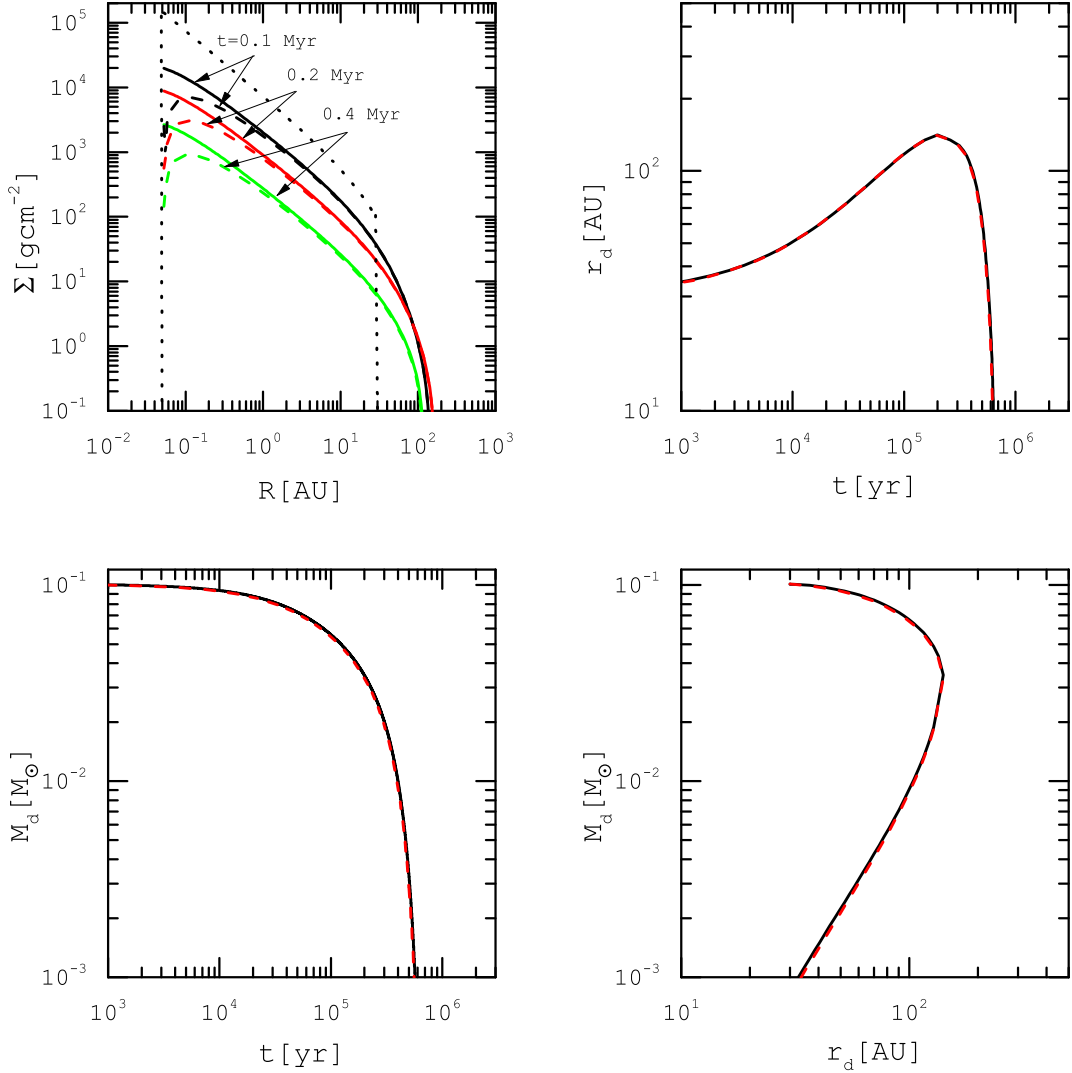}
\caption{On the top left-hand panel, the surface density profile of a circumstellar disk with winds driven by  external FUV radiation is shown at different times, as labeled. The solid and dashed curves correspond to the solutions with the BCs of VGR2016 and AAC2013, respectively. The initial surface density distribution is shown by a dotted curve. The adopted input parameters are $s=1/2$, $\alpha=0.01$, $r_{\rm in}=0.05$ AU. On the top right-hand panel, disk radius as a function of time is shown. On the bottom row, profiles of the disk mass (left) and evolutionary track in the plane of disk mass and radius (right) are shown for the presented solutions.}\label{fig:f1}
\end{figure*}

The third key quantity is the wind mass-loss rate, i.e., $\dot{\Sigma}_{\rm w}$. Its mechanism is commonly attributed to the magnetically or photoevaporative mechanisms. When a disk is exposed to a strong radiation field, this photoevaporative process leads to a significant disk mass removal. Source of radiation, however, is either the host star or ambient radiation field. When the photoevaporative wind is driven by the host star radiation, the mass-loss rate is parameterized as follows \citep{Hollenbach1994} 
\begin{equation}\label{eq:internal}
\dot{\Sigma}_{\rm w}=\frac{\dot{M}_{\rm wind}}{4\pi {r_{c}}^{2}}(\frac{r}{r_{c}})^{-5/2}, r\geq r_{c}
\end{equation}
where $\dot{M}_{\rm wind}$ is the integrated mass-loss rate. For X-ray photoevaporation, we can adopt these values: $\dot{M}_{\rm wind}=1.0\times 10^{-8}$  M$_{\odot}$yr$^{-1}$ and $r_c =5$ AU  \citep{Owen2010, Owen2011}.

Following models of \cite{Adams2004} and AAC2013, the mass-loss rate due to FUV photoevaporation by the external stars is written as
\begin{equation}\label{eq:external}
\dot{\Sigma}_{\rm w}=\frac{C n_{\rm d} \sqrt{k_{\rm B} T_{\rm ph}\mu}}{4\pi}\left ( \frac{r_{g}}{r} \right )^{\frac{3}{2}}\left [1+\frac{r_{g}}{r}\right ] \exp \left (-\frac{r_{g}}{2r} \right ),
\end{equation}
where $C$ is a constant of order unity and $n_{d} \approx 10^3-10^8 {\rm {cm}}^{-3}$ is the gas number density at the disk outer edge \citep{Adams2004}. The critical radius $r_{g}$ is defined as a radial distance where the sound speed is comparable to the escape velocity and $T_{\rm ph}$ is the heated gas temperature at the critical radius. Although location of $r_{g}$ strongly depends on the radiation field flux, we have an analytical relation for the critical radius \citep{Adams2004}, 
\begin{equation}\label{eq:Tph}
r_{g}=\frac{G M_{\rm c}\mu}{k_{\rm B} T_{\rm ph}}\approx 226{\rm {AU}} (\frac{M_{\rm c}}{M_{\odot}}) (\frac{T_{\rm ph}}{1000 {\rm K}})^{-1}. 
\end{equation}
According to detailed models \citep{Adams2004}, the critical radius becomes $r_{g} = 157$ AU when a disk is exposed to FUV flux level $G_{0} = 3000$. Note that typical flux level in the interstellar medium corresponds to $G_{0} =1$. The critical radius $r_{g}=157$ AU corresponds to heated gas temperature $T_{\rm ph} \approx 1440 {\rm K}$ and the number density $n_d\approx10^{6} {\rm {cm}}^{-3}$. In our analysis, these values are adopted as canonical values unless otherwise is stated. In Table \ref{tab:1} we list our model quantities and their symbols.

\section{analysis}
We solve the surface density evolution equation (\ref{eq:main}) subject to the boundary conditions (BCs) implemented by VGR2016 and AAC2013 (see Eqs. (\ref{eq:VGR}) and (\ref{eq:ANN})). An implicit finite difference method is adopted that are distributed logarithmically. A large outer boundary is adopted, i.e. $r_{\rm out}=20000 $ AU to ensure that the solutions are independent of the outer edge. The inner edge $r_{\rm in}$ of a circumbinary disk, however, is chosen depending on semimajor axis $a_{\rm b}$. We also verified that our solutions are consistent with previous studies  AAC2013 and VGR2016.   
  
Following most previous studies \citep[e.g.,][]{Alexander12, Martin13}, the initial surface density distribution within the range $r_{\rm in}<r<r_{\rm d0} $ is given by an exponentially truncated profile \citep[e.g.,][]{Lyn74}
\begin{equation}\label{eq:surf-ini}
\Sigma (r,0)=\frac{M_{\rm d0}}{2\pi \left [ \exp(-\frac{r_{\rm in}}{r_{\rm d0}})-e^{-1} \right ] r_{\rm d0} r} \exp (-r/r_{\rm d0}),
\end{equation}
and $\Sigma (r,0)=0$ for $r<r_{\rm in}$ or $r>r_{\rm d0}$. Here, $r_{\rm d0}$ and $M_{\rm d0}$ are the initial radius and mass of the disk, respectively. The initial disk mass is set to $M_{\rm d0}=0.1$ $M_{\rm c}$ to focus on exploring role of the other model parameters. Furthermore, the initial disk size is $r_{\rm d0}=30$ AU. Note that our adopted initial surface density distribution (\ref{eq:surf-ini}) is slightly different from \cite{Alexander12} to ensure that the initial disk mass is $M_{\rm d0}=0.1$ $M_{\rm c}$. But the initial surface density of \cite{Alexander12} corresponds to the disks with a smaller initial mass and thereby a shorter lifetime.

\begin{figure*}
\includegraphics[scale=1.0]{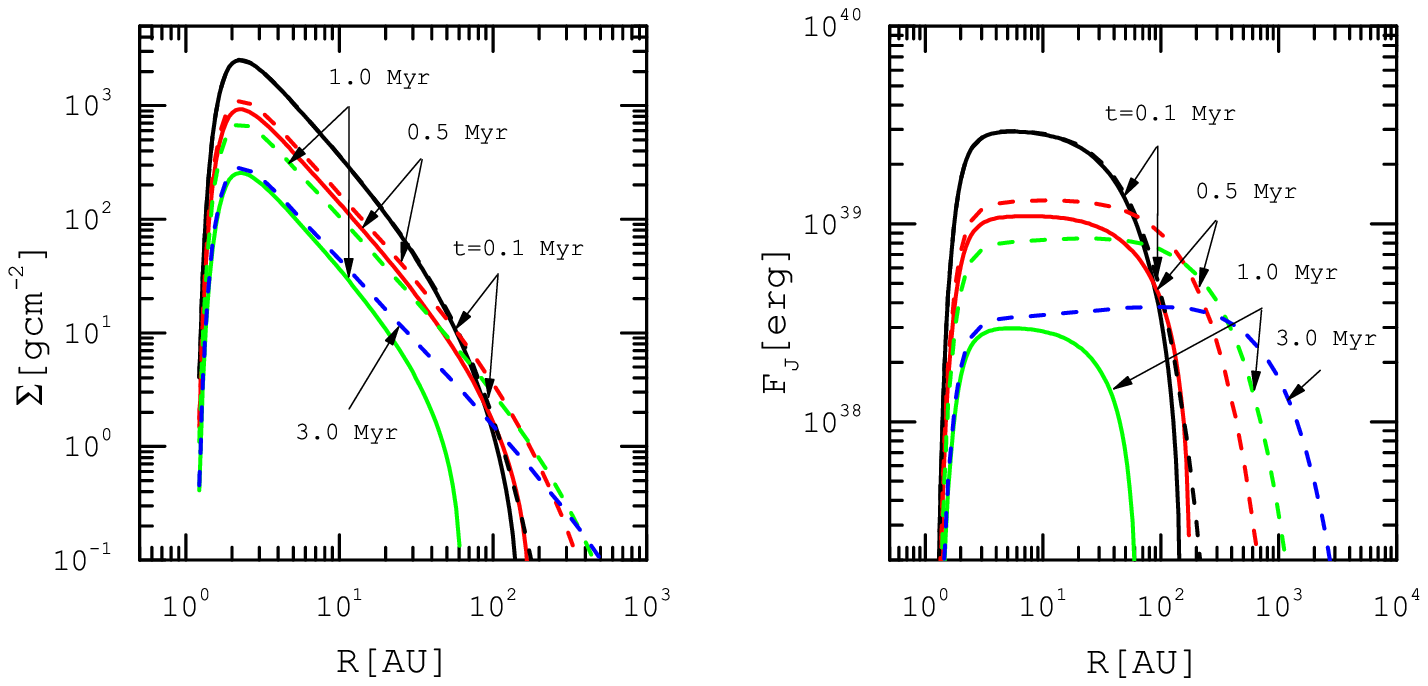}
\caption{Profiles of the surface density (left) and the viscous angular momentum flux (right) for a circumbinary disk at different times, as labeled. The circumbinary disk evolution with winds are displayed by solid curves, whereas solutions without winds are shown by dashed curves. Masses of the primary and secondary stars are $M_{\rm p}=M_{\rm s}=0.5$ M$_{\odot}$ and their separation is taken to be $a_{\rm b}=0.2$ AU. Other model parameters are $s=1/2$, $f=1$, $\alpha=0.01$, $G_0=3000$, $r_g=157$ AU, and $n_d=10^6{\rm{cm}^{-3}}$. }\label{fig:f2}
\end{figure*}

VGR2016 introduced the following BCs for the circumstellar disks:
\begin{equation}\label{eq:VGR}
\frac{\partial F_{J}}{\partial l}|_{\rm r_{\rm in}}=\frac{F_{J}(r_{\rm in})}{l_{\rm in}}, ~\frac{\partial F_{J}}{\partial l}|_{\rm r_{\rm out}}=0,
\end{equation}
where the viscous angular momentum flux $F_J$ is defined as $F_J =3\pi\nu\Sigma l$ and $l=\Omega r^2$ is the specific angular momentum. Note that  AAC2013 used the standard BCs, where the surface density tends to zero at the inner edge and the disk can expand freely at the outer radius. We note that binary torque at the inner edge reduces the accretion rate. Nevertheless, the adopted BCs by \cite{Martin13} in a circumbinary PPD can be written as  
\begin{equation}\label{eq:ANN}
\Sigma (r,t)_{\rm r_{\rm in}}=0, \frac{\partial F_{J}}{\partial l}|_{\rm r_{\rm out}}=0.
\end{equation}
The profile of $F_{J}$ in a circumstellar disk rapidly converges to $r^{1/2}$, whereas in a circumbinary disk $F_J$ tends to a flat distribution due to the exerted binary torque at the inner edge. This interesting feature, however, was found in the absence of the winds (VGR2016). Whether or not this feature is preserved in a circumbinary disk with the photoevaporative winds is examined in Figure \ref{fig:f2}. 

\begin{figure*}
\includegraphics[scale=0.95]{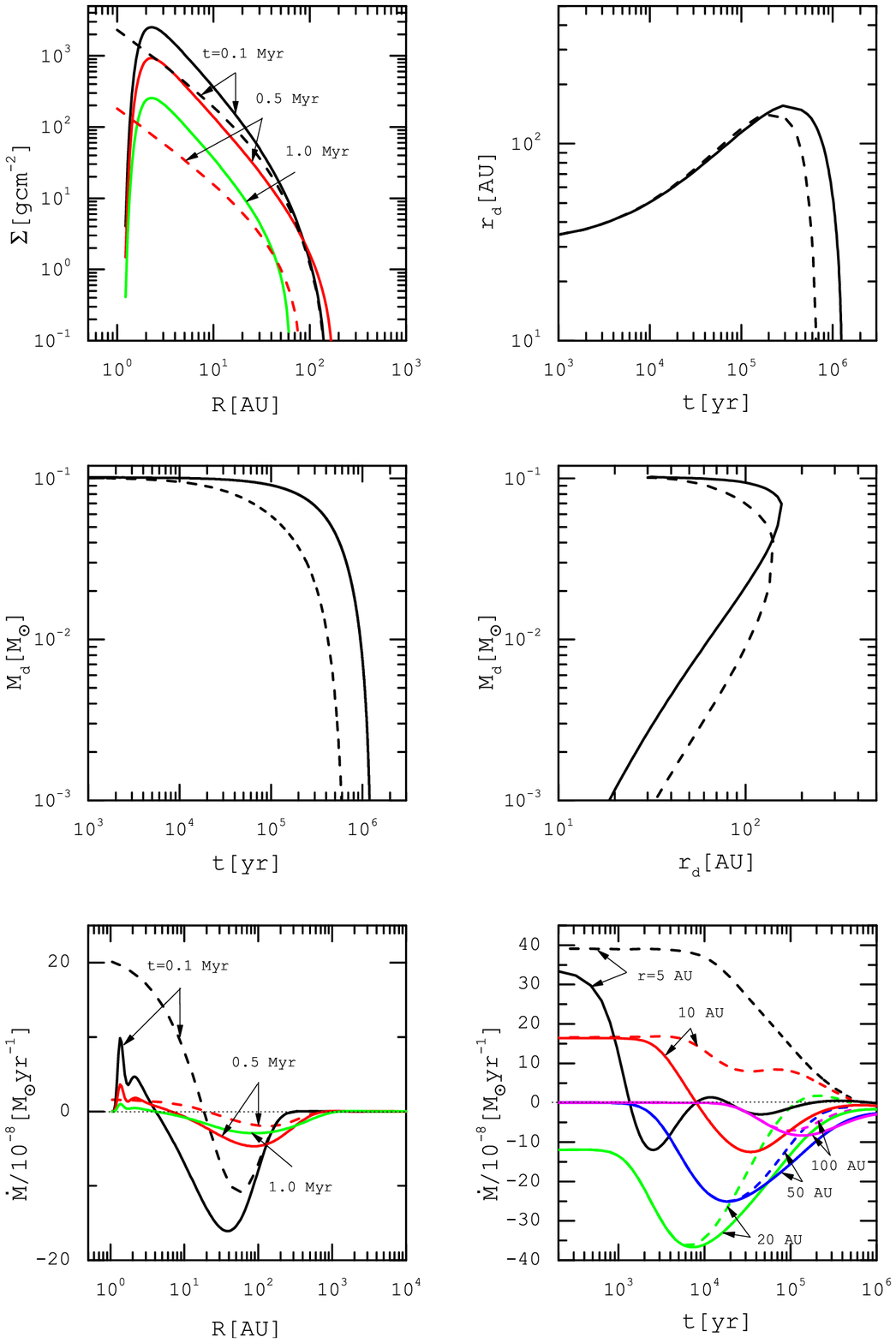}
\caption{Physical quantities of a circumbinary disk (solid curves) and an identical single-star disk (dashed curves). The mass-loss is driven by external FUV radiation field. The total mass of the binary is $M_{\rm c}=1$ M$_\odot$ and the separation of its components is assumed to be $a_{\rm b}=0.2$ AU. Other model parameters are $s=1/2$, $q=1.0$, $f=1.0$, $\alpha=0.01$, $G_0=3000$, and $r_g=157$ AU. On the top left-hand panel, surface density distribution is shown at different epochs, as labeled. On the top right-hand panel, disk radius is shown through time. On the middle left-hand panel, disk mass as a function of time is shown. On the middle right-hand panel, the locus of points in the plane of disk mass and radius corresponding to the explored cases are shown. The accretion rate as a function of the radial distance (the left-hand panel) and of time (the right-hand panel) are shown on the bottom panels. Note that positive values of the accretion rate correspond to the radial motion of the gas toward the central object.}\label{fig:f3}
\end{figure*}

To illustrate differences in a circumstellar disc evolution subject to the BCs adopted by AAC2013 and VGR2015, we first perform evolutionary calculations for the circumstellar disks using these BCs. Figure \ref{fig:f1} displays quantities of a circumstellar disk exposed to external FUV radiation field with the BCs used by VGR2016 (solid curves) and AAC2013 (dashed curves). In this figure, we adopt $r_{\rm in}=0.05$ AU and use a logarithmically spaced radial grid with 258 cells. The host star mass is fixed at 1 M$_{\odot}$. The initial disk mass and radius are $M_{\rm d0}=0.1 M_{\odot}$ and $r_{\rm d0}=30$ AU, respectively. Other model parameters are $s=1/2$, $\alpha=0.01$, $G_0=3000$, $r_g =157$ AU, and $n_d=10^6{\rm{cm}^{-3}}$. Evolution of the disk surface density at different times, as labeled, is shown on the top left-hand panel of this figure. The initial surface density distribution is shown by a dotted curve too. The surface density with the BCs used by VGR2016 is slightly larger than the surface density subject to the BCs implemented by AAC2013.

To explore evolution of the disk size, inspired by  AAC2013, we define disk radius $r_d$ as a radius where the enclosed disk mass $M_{\rm enc}$ is a given fraction $\chi$, say 0.99, of the total disk mass $M_{\rm d}$. Thus, we have 
\begin{equation}\label{eq:rd}
M_{\rm enc}(r_d,t)=\chi M_d(t).
\end{equation}
With a lower $\chi$, a disk has a smaller size, however, AAC2013 showed that evolution of the disk radius $r_d$ does not depend on the adopted fraction $\chi$. We, therefore, define a disk radius with $\chi=0.99$. On the top right-hand panel of Figure \ref{fig:f1}, disk radius evolution is shown. The disk size gradually increases due to the viscous stress, however, the evaporative wind is very efficient in mass removal in the outer region. Therefore, disk shrinking is started after about $10^5$ years due to the photoevaporative wind. On the bottom left-hand panel of Figure \ref{fig:f1}, disk mass as a function of time is shown. While the total mass of the disk does not show a noticeable reduction at early phases of the disk evolution, photoevaporative wind eventually becomes effective in mass removal from the disk. We note that disk mass profile is a decreasing function of the time irrespective of the disk size. We now define disk lifetime as a time period that total disk mass reduces to one percent of its initial mass. In the light of this definition, disk lifetime is about $0.6$ Myr irrespective of the adopted BCs. We also find that profiles of the disk mass and radius are almost independent of the imposed BCs. On the bottom right-hand panel of Figure \ref{fig:f1}, evolutionary track in the plane of disk mass and radius are shown.

Figure \ref{fig:f2} depicts the surface density distribution (left) and the corresponding $F_J$ profile (right) for a circumbinary disk with photoevaporative wind due to external FUV radiation (solid curve) and without wind (dashed curve). Different colors correspond to different epochs, as labeled. The binary mass ratio is $q=1$ and its total mass is $M_{\rm c}=1$ M$_{\odot}$. The binary separation is assumed to be $a_{\rm b}=0.2$ AU and the inner edge is taken to be $r_{\rm in}\simeq 5 a_{\rm b}=1.0$ AU. Other model parameters are $\alpha=0.01$, $G_0=3000$, $r_g = 157$ AU, $f=1.0$, $n_d=10^6{\rm{cm}^{-3}}$ and $s=1/2$. Although reduction of the surface density with time is due to the viscous stress, this reduction is more significant when photoevaporative wind induced by the external radiation source is considered. Disk spreading in the absence of the wind is more evident, however, photoevaporative winds strongly deplete outer regions and create a sharp outer edge. As time proceeds, surface density reduction in the case with a wind becomes more significant in comparison to the no-wind solution. We also find that the angular momentum flux of a circumbinary disk is more or less a flat distribution irrespective of the wind presence. 

Figure \ref{fig:f3} provides a comparison between physical quantities of a circumbinary disk (solid curve) and a circumstellar disk counterpart (dashed curve) in the presence of winds induced by external FUV radiation field with the flux level $G_0=3000$. An equal-mass binary with separation $a_{\rm b}=0.2$ AU and the total mass $M_{\rm c}=1.0$ M$_{\odot}$ is considered. In the case of a circumstellar analogous, a single solar mass star at the disk center is considered and the adopted BC is VGR2016. As before, the temperature exponent, torque parameter and viscosity coefficient are $s=1/2$, $f=1.0$ and $\alpha=0.01$, respectively. The initial surface density distribution is given by equation (\ref{eq:surf-ini}) and its evolution for different times, as marked, is shown in the top left-hand panel of Figure \ref{fig:f3}. During the early phase of the evolution, disk surface density in its outer part does not change, however, the accretion from this region gradually piles up in the inner region due to the binary torque. This accumulated mass, thereby, is transferred outward where the photoevaporative wind plays a crucial role in mass removal. The circumbinary disk, for instance, evolves over 0.5 Myr with a surface density by up to a factor of 10 larger than a circumstellar disk analogous. During about one million years, the circumstellar disk is cleared entirely, whereas the circumbinary gas disk is still present. This trend is a direct consequence of the binary torque that leads to gas pileup in the innermost region and viscous spreading of the accumulated gas.

The top right-hand panel of Figure \ref{fig:f3} plots disk radius versus time. As before, the disk radius is defined using equation (\ref{eq:rd}). Disk radius increases at early times no matter a binary or a single star resides at the disk center. As time proceeds, however, further growth of the disk size is prevented due to the efficient mass removal by the photoevaporative winds. Thereafter, disk radius rapidly decreases  with time. The circumbinary disk extends to the maximum size 156 AU in about 0.3 Myr, whereas the maximum size of a  similar circumstellar disk is 141 AU during 0.2 Myr. 

On the middle left-hand panel, disk mass as a function of time is shown. It is a decreasing function of time irrespective of the disk size and the existence of a binary or a single star at the center. Photoevaporative wind as a mass-loss mechanism leads to this obvious feature. We have already defined a disk lifetime as the time by which  disk losses 99 percent of its initial mass. We, therefore, find that the circumbinary disk survives over about 1.2 Myr which is by a factor of two longer than a circumstellar disk counterpart. In other words, circumstellar disk mass declines faster compared to the circumbinary case. Corresponding tracks in the plane of disk mass and radius are also shown in the middle right-hand panel of Figure \ref{fig:f3}. 

On the bottom left-hand panel of Figure \ref{fig:f3}, the accretion rate versus the radial distance is shown for different times, as labeled. Note that for the mass inflow  toward the center of the system, the accretion rate is positive. The solid curve corresponds to the circumbinary case, whereas dashed curves represent an identical single-star disk. There is always mass inflow at the inner regions in both circumstellar and circumbinary disks, however, the accretion rate at the inner edge of a circumbinary disk tends to zero due to the binary torque. But in the circumstellar case, the mass accretion rate at the inner edge is significant. We generally find that the accretion rate in the circumbinary disk is smaller than that in the circumstellar case. Beyond a certain radial distance, however, there is mass flow toward outer disk edge. Size of the region with mass inflow toward the center of system is smaller in the circumbinary case in comparison to the circumstellar disk. As time proceeds, therefore, a circumstellar disk quickly losses its mass because of its high accretion rate onto the central star and relatively small amount of mass is lost due to the wind mass removal in the outer region. In contrast, accretion rate onto the binary is suppressed due to the binary torque and the mass pileup  is transfered to the outer region where photoevaporative winds are efficient and hence disk lifetime is determined by the mass-loss rate in this region. On the bottom right-hand panel, we display evolution of the accretion rate at several different locations in the disk, as labeled. It also shows that the accretion rate at a given radial distance, say 5 AU, is strongly suppressed in the circumbinary disks in comparison to the circumstellar disks. At large radial distance, however, there is mass flow toward outer disk edge and a negligible discrepancy between the mass accretion rate in the circumstellar and circumbinary disks is found due to weakness of the binary torque.  

\begin{figure}
\includegraphics[scale=1.0]{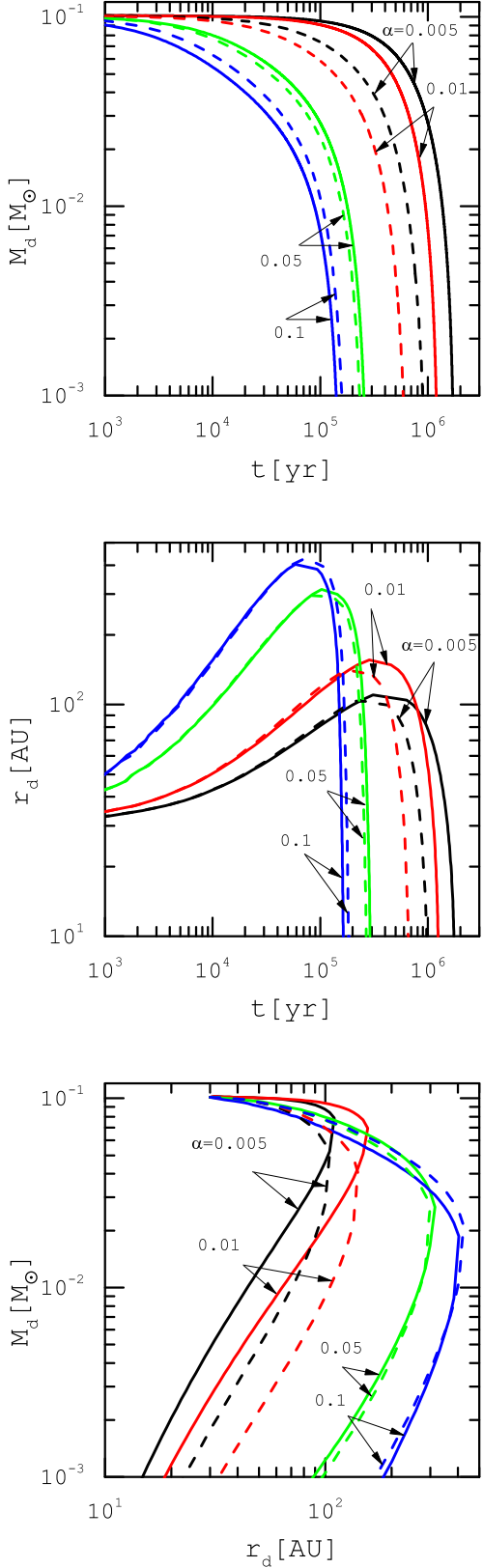}
\caption{The disk mass (top), its radius (middle) and the corresponding track in the disk mass and radius plane (bottom) are shown for different values of the viscosity parameter, as labeled. The input parameters are $s=1/2$, $a_{\rm b}=0.2$ AU, $q=1.0$, $f=1.0$, and $G_0=3000$. The solid and dashed curves correspond to the circumbinary disk and a similar circumstellar disk, respectively.}\label{fig:f4}
\end{figure}

Spreading of the disk due to the viscous torque occurs over the viscous timescale which depends on the viscosity parameter $\alpha$. In Figure \ref{fig:f4}, we explore the role of the viscosity coefficient in the evolution of a circumbinary disk (solid curve) and a circumstellar disk counterpart (dashed curve) by considering different values of this parameter within the interval $0.005\leq \alpha \leq 0.1$, as labeled. We consider an equal mass binary with the total mass $M_{\rm c} =1$ M$_\odot$ and the rest of model parameters are $G_0=3000$, $a_{\rm b}=0.2$ AU, and $q=f=1.0$. On the top panel, disk mass as a function of time is shown. We find that disk lifetime strongly depends on the adopted viscosity coefficient. A disk can survive over a longer time period if a lower viscosity coefficient is considered. A key feature of the viscous disk evolution is its extension to the larger radii due to the angular momentum transport. A higher viscosity coefficient, therefore, leads to a faster redistribution disk material. Photoevaporative winds, on the other hand, are more efficient in the disk outer regions. Mass removal, therefore, in a disk with a large viscosity coefficient is more efficient, which leads to the disk dispersal over a shorter time. Disk mass evolution in the circumbinary and circumstellar cases with a large viscosity coefficient, say $\alpha =0.1$, does not seem to be different. As the viscosity tends to the lower values, however, a discrepancy between the temporal evolution of the total disk mass in the circumbinary and circumstellar cases becomes  more pronounced. In the cases with $\alpha=0.01$, for instance, the lifetime of a circumbinary PPD is about 1.2 Myr, whereas a circumstellar disk counterpart is dispersed over a shorter time, i.e. 0.6 Myr. Note that AAC2013 obtained a shorter lifetime for a single star disk with $\alpha=0.01$ and $G_0=3000$, i.e. about 0.2 Myr. This discrepancy is probably due to adopting different model parameters. We find that for $\alpha=0.1$, lifetimes of the circumbinary and circumstellar PPDs are about 0.15 Myr. A comparison between cases with $\alpha=0.1$ and 0.01 shows that a circumbinary PPD lifetime is enhanced by a factor of eight, whereas in the circumstellar case this lifetime enhancement is about a factor of four. We, therefore, find that $\alpha -$dependence of the disk lifetime in the circumbinary case is stronger than the circumstellar case. 

The middle panel of Figure \ref{fig:f4} shows the temporal evolution of the disk radius for different values of the viscosity coefficient, as labeled. A disk extends out at the early phase of its evolution to reach the largest size where subsequent rapid disk shrinking is started due to the  photoevaporative wind. However, the maximum attainable disk size reduces with the viscosity coefficient. A circumbinary disk, for instance, can spread out with time to about 400 AU for $\alpha=0.1$, whereas the maximum disk size is about 150 AU for $\alpha=0.01$. A similar trend is found for the circumstellar disks too. In the bottom panel of Figure \ref{fig:f4}, the corresponding tracks in the disk mass and radius plane are shown for different viscosity parameters.  

\begin{figure}
\includegraphics[scale=1.0]{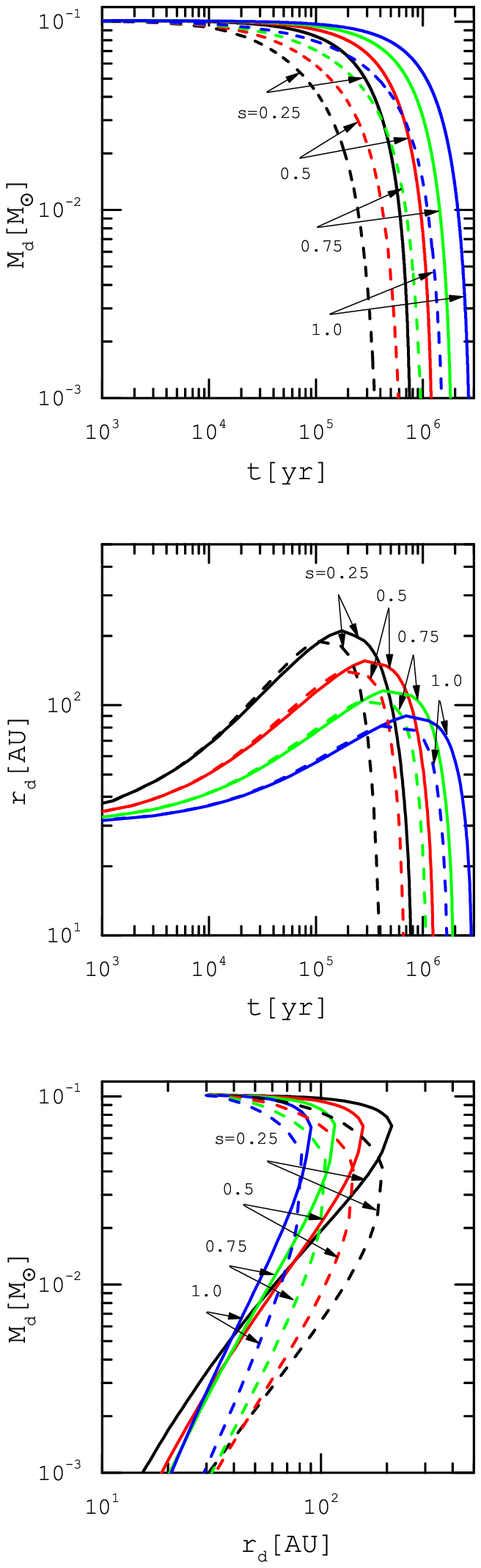}
\caption{Similar to Figure \ref{fig:f4}, but for $\alpha=0.01$ and different values of the temperature exponent $s$, as labeled. The solid and dashed curves correspond to a circumbinary disk and a similar circumstellar disk, respectively. }\label{fig:f5}
\end{figure}

The other key model parameter is the temperature exponent $s$ where its role in the evolution of the circumbinary and circumstellar PPDs is examined in Figure \ref{fig:f5}. All model parameters are similar to Figure \ref{fig:f4}, except the temperature exponent with different values, as labeled. In Figure \ref{fig:f5}, the viscosity parameter is 0.01. We can infer disk lifetime from the top panel which shows disk mass evolution. When the temperature exponent is $s=1.0$, lifetimes of the circumbinary and circumstellar PPDs are found about 2.7 Myr and 1.5 Myr, respectively. But these lifetimes are reduced to 0.75 Myr and 0.35 Myr for $s=0.25$. It shows that disk lifetime increases as the temperature distribution becomes steeper. This behavior is understood in terms of the viscosity which controls disk spreading rate. Since viscosity in the $\alpha -$model is directly proportional to the disk temperature, it becomes a steep function of the radial distance and tends to low values when the temperature distribution is steep. We have already found that disk spreading due to the viscous torque is slower in the case with a low viscosity which leads to an enhancement of the disk lifetime.

\begin{figure}
\includegraphics[scale=1.0]{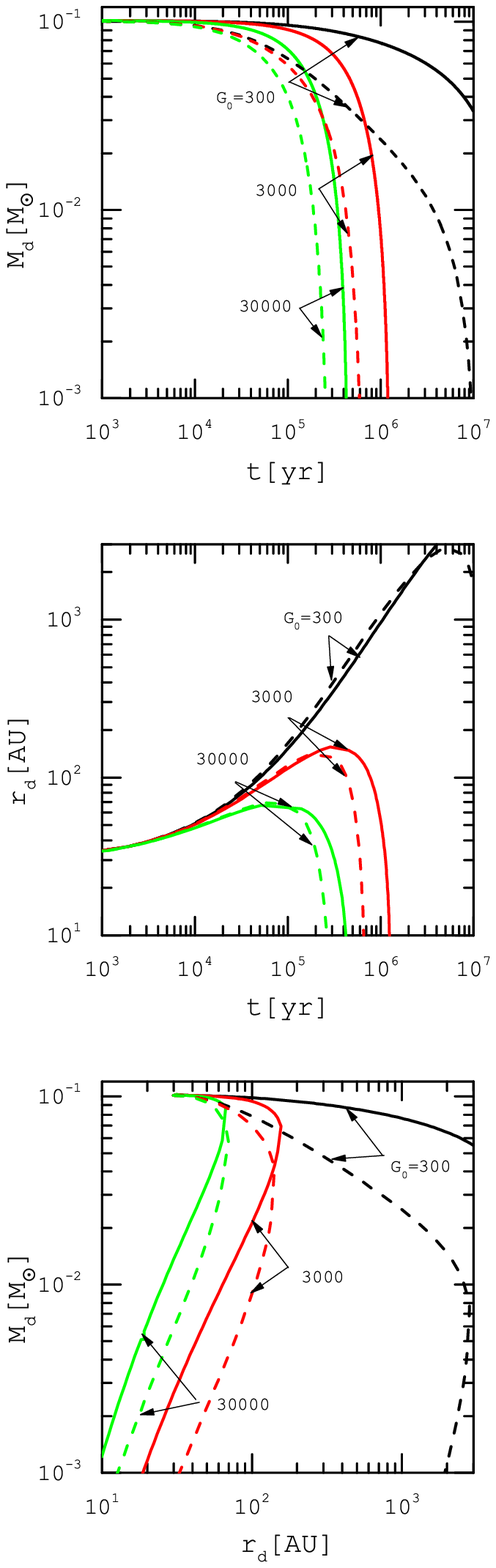}
\caption{Similar to Figure \ref{fig:f4}, but for $\alpha=0.01$ and different values of $G_0$, as labeled. The solid and dashed curves correspond to a circumbinary disk and a similar circumstellar disk, respectively. }\label{fig:f6}
\end{figure}

\begin{figure}
\includegraphics[scale=1.0]{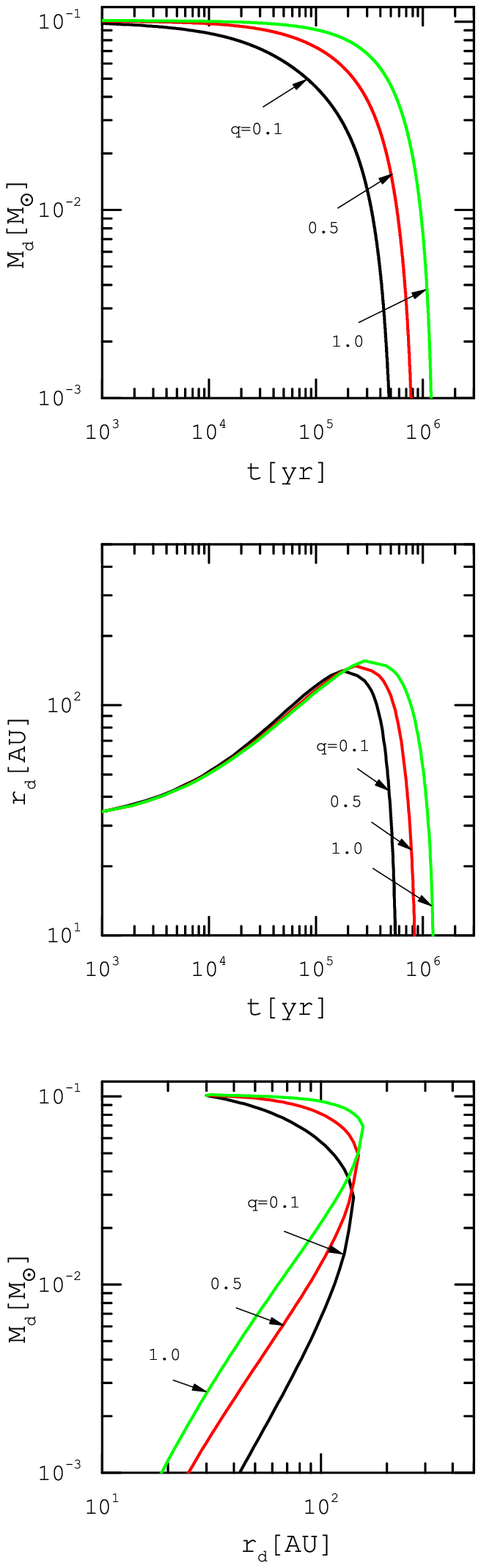}
\caption{The circumbinary disk mass (top), its radius (middle) and the corresponding track in the disk mass and radius plane (bottom) are shown for different values of the binary mass ratio, as labeled. As before, the binary total mass is $M_{\rm c} =1.0$ M$_{\odot}$ and the other model parameters are $f=1.0$, $a_{\rm b}=0.2$ AU, $G_0=3000$, $s=1/2$, and $\alpha=0.01$.}\label{fig:f7}
\end{figure}

\begin{figure}
\includegraphics[scale=1.0]{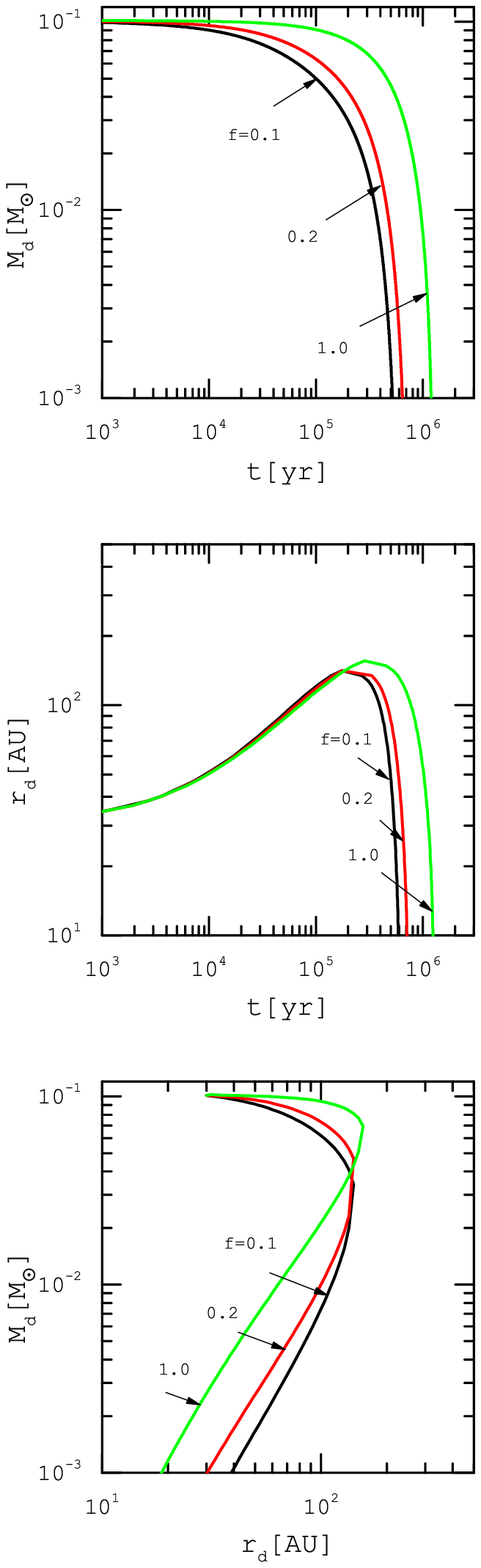}
\caption{Similar to Figure \ref{fig:f7}, but for equal-mass binaries and different values of the binary torque coefficient $f$, as labeled. }\label{fig:f8} 
\end{figure}

\begin{figure}
\includegraphics[scale=1.0]{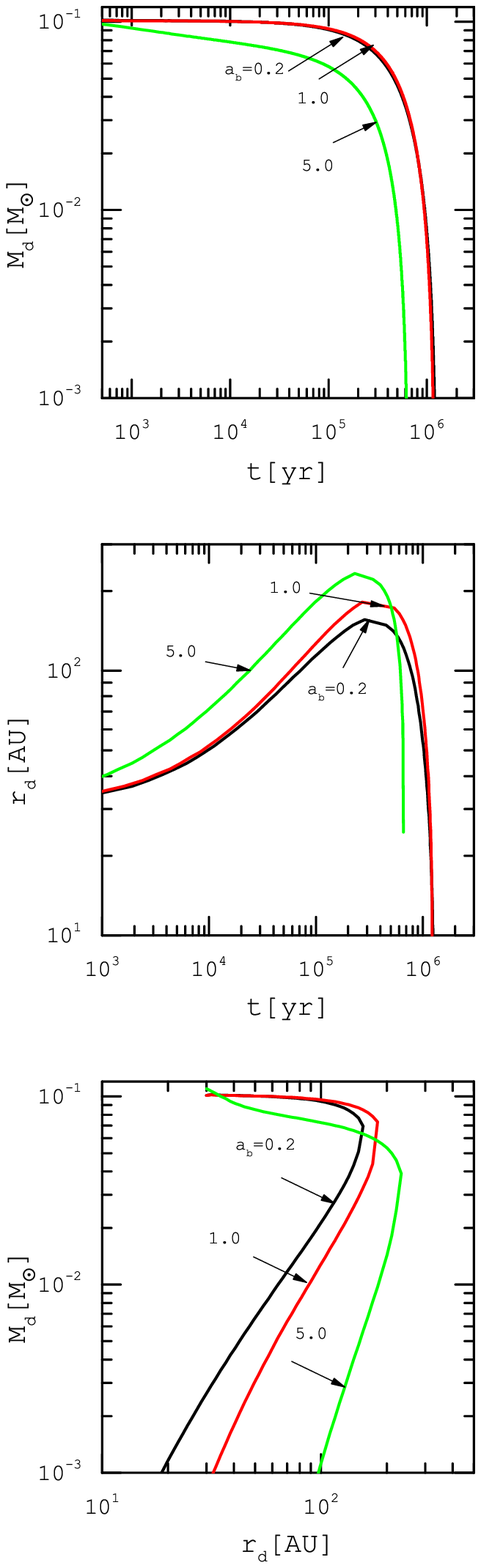}
\caption{Similar to Figure \ref{fig:f7}, but for equal-mass binaries and different values of the semimajor axis $a_{\rm b}$, as labeled. }\label{fig:f9} 
\end{figure}

The middle panel of Figure \ref{fig:f5} exhibits disk size evolution with time for different temperature exponents, as labeled. While a circumbinary disk with $s=1.0$ attains a maximum size 90 AU during 2.7 Myr, for $s=0.25$  the disk reaches to the largest size 210 AU during 0.75 Myr. Therefore, the maximum size of a circumbinary disk is larger as the temperature exponent decreases. We also find a similar trend for the circumstellar disks. For a given $s$, however, the maximum size of a circumbinary disk is found slightly larger than a similar circumstellar disk. The bottom panel of Figure \ref{fig:f5} displays corresponding tracks in the plane of the disk mass and radius for different values of the temperature exponent.

In Figure \ref{fig:f6}, we explore the influence of FUV flux level $G_0$ in the evolution of the disk mass (top), disk size (middle) and the corresponding tracks in the disk mass and radius plane (bottom) for disks with the viscosity parameter $\alpha=0.01$. We suppose that the flux level can vary from 300 to 30000 and the other parameters are adopted similar to Figure \ref{fig:f4}. The solid and dashed curves correspond to the circumbinary and circumstellar disks, respectively. If we assume that $G_0 =30000$, the critical radius becomes $r_g =90 $ AU and the equation \ref{eq:Tph} yields $T_{\rm ph}=2510$ K which is associated to $n_d =10^7 {\rm {cm}}^{-3}$ \citep{Adams2004}. For a radiation field strength $G_0=300$, we find that $r_g = 357$ AU and $T_{\rm ph}=633$ K. This value of $T_{\rm ph}$ corresponds to $n_d =10^3 {\rm {cm}}^{-3}$.  Note that for the flux level $G_0=3000$ we have $r_g=157$ AU, $T_{\rm ph}=1440$ K, and $n_d =10^6 {\rm {cm}}^{-3}$. We find that the lifetimes of circumbinary disks exposed to FUV flux level $G_0=30000$ is about 0.42 Myr, whereas the circumstellar disk subject to the same FUV flux level   can survive over 0.25 Myr. For FUV flux level $G_0 =300$, the lifetime of circumstellar disk is about 9.4 Myr, whereas, the circumbinary disks exposed to FUV flux level $G_0=300$ are not dispersed after 10 Myr. Therefore, the photoevaporative wind due to FUV flux level $G_0 =300$ is not efficient mass removal mechanism. Furthermore, size of  the circumbinary disks  in the presence of winds with flux level $G_0=30000$ does not exceed 70 AU, but when the external FUV flux level is reduced to  $G_0=300$, disk size may extend to several thousand AU. 

Figure \ref{fig:f7} shows the role of the mass ratio $q$ in the evolution of the disk mass (top), disk size (middle) and the corresponding tracks in the disk mass and radius plane (bottom). The total binary mass is fixed at $M_{\rm c} =1.0$ M$_{\odot}$ and the temperature exponent is $s=1/2$, but different mass ratios are considered, as labeled. Other model parameters are similar to Figure \ref{fig:f5}. We find that evolution of a disk with a high binary mass ratio is slower because the binary torque which causes  mass pileup in the innermost region is directly proportional to the mass ratio. While a circumbinary disk with an extreme mass ration $q=1.0$ survives over 1.2 Myr, a similar disk with a smaller mass ratio $q=0.1$ is depleted during about 0.5 Myr. 

In our adopted relation for the binary torque, i.e. Eq. (\ref{eq:torque}), the parameter $f$ with a value from 0.001 to 1 was introduced. We now examine to what extent our solutions are dependent on the adopted value of the binary torque coefficient $f$. Figure \ref{fig:f8} shows disk mass (top), radius (middle) and the corresponding tracks in the disk mass and radius plane (bottom) for $M_c =1.0$ M$_{\odot}$ and $q=1.0$ and different values of the binary torque coefficient $f$, as labeled. As the parameter $f$ reduces, the binary torque becomes weaker and disk dispersal happens during a shorter time. For instance, the lifetime of a disk with $f=1.0$ is 1.2 Myr, whereas, for $f=0.1$, the disk lifetime reduces to about 0.52 Myr. 

The effect of the semimajor axis in the evolution of the disks around an equal-mass binary is explored in Figure \ref{fig:f9}. In this figure, the semimajor axis $a_{\rm b}$ is assumed to be 0.2 AU, 1.0 AU, and 5.0 AU, as labeled. Note that, for example, the binary systems of AK Sco, DQ Tau, and GG Tau Ab have the semimajor axis of order 0.16 AU \citep{Alencar2003}, 0.13 AU \citep{Czekala2016}, and 4.5 AU \citep{diFolco2014}, respectively. Since we have supposed that the inner radius is $5 a_{\rm b}$ and the initial disk size is 30 AU, we cannot consider larger values of $a_{\rm b}$. Other model parameters are similar to Figure \ref{fig:f7}. We find that circumbinary disks with $a_{\rm b}=1.0$ AU have a lifetime of order 1.1 Myr slightly shorter than the lifetime of circumbinary disks with $a_{\rm b}=0.2$ AU. When the semimajor axis is assumed to be 5.0 AU, disk lifetime decreases by a factor of 2 compared to disk with $a_b=0.2$ AU, i.e., 0.62 Myr. At this time, such a disk with $a_{\rm b}=5$ AU shrinks to radii $r_d\leq 100$ AU. However, the disks with $a_{\rm b}=0.2$ AU and $a_{\rm b}=1.0$ AU shrink to $r_d\leq 19$ AU and $r_d\leq 33$ AU, respectively. The disk with $a_{\rm b}=5.0$ AU may extend to 233 AU, whereas the maximum size of a disk with $a_{\rm b}=0.2$ AU is about 156 AU.  These trends are understandable in terms of the binary torque scaling with the semimajor axis.  The binary torque equation (\ref{eq:torque}) shows that the torque becomes weaker and the resulting mass pileup becomes less significant with increasing the semimajor axis. Viscous disk spreading, on the other hand, is still efficient in extending the disk size to regions where mass-loss rate by phtoevaporative winds is large. In a circumbinary disk with a large semimajor axis, therefore, mass removal by winds becomes more efficient in comparison to a similar disk with a small semimajor axis. It then reduces disk lifetime with increasing the semimajor axis.

\begin{figure*}
\includegraphics[scale=1.0]{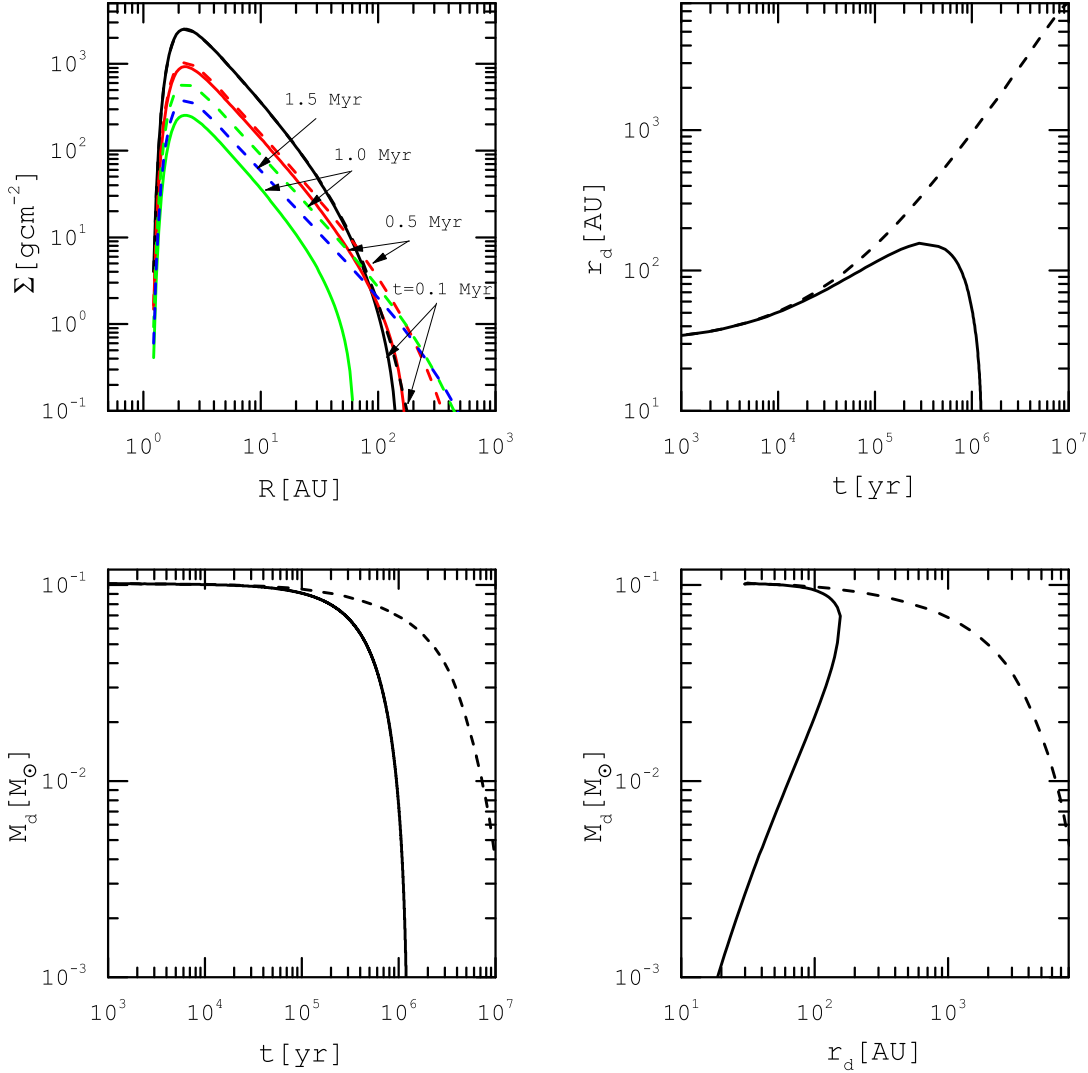}
\caption{On the top left-hand panel, the surface density profile of a disk orbiting an equal-mass binary with a total mass $M_{\rm c}=1$ M$_{\odot}$ is shown for different times, as labeled. The solid and dashed curves correspond to the cases with external FUV radiation driven wind ($G_0=3000$) and the host star X-ray driven wind, respectively. The input parameters are $f=1.0$, $a_{\rm b}=0.2$ AU, $s=1/2$ and $\alpha=0.01$. On the top right-hand panel, disk radius as a function of time is shown. On the bottom row, profiles of the disk mass (left) and evolutionary tracks in the plane of disk mass and radius (right) are shown for the presented solutions.}\label{fig:f10}
\end{figure*}

The source of photoevaporative radiation, as mentioned before, can be either the host star or ambient stars. We have so far incorporated external FUV photoevaporation with the mass removal rate parameterized by equation (\ref{eq:external}) \citep{Adams2004}. We can now explore the relative importance of the photoevaporative winds driven by the host star and the external FUV radiation fields. Equation (\ref{eq:internal}) gives photoevaporative rate due to the host star X-ray radiation \citep{Hollenbach1994}. Figure \ref{fig:f10} provides this comparison by considering disks with external FUV photoevaporative winds (solid curve) and  photoevaporative winds due to the host star X-ray radiation (dashed curve). The model parameters in either of these two cases are $M_{\rm c}=1.0$ M$_{\odot}$, $q=1.0$, $f=1.0$,  $a_{\rm b}=0.2$ AU, $s=1/2$ and $\alpha =0.01$. The evolution of the disk surface density at different times, as labeled, is shown in the top left-hand panel of this figure. 

Mass removal by the externally FUV radiation induced wind affects outer disk region even at the early times, whereas this effect is not efficient in the disk with wind driven by the host star X-ray radiation. As time proceeds, however, the effect of the wind mass-loss gradually appears at smaller radii. On the top right-hand panel, disk radius is shown for the explored cases. The disk radius gradually increases in a case with external FUV radiation driven wind to reach a maximum size where this trend is suppressed due to the efficient mass removal. In the case of wind driven by the host star X-ray radiation, however, mass-loss is not strong enough to stop the viscous spreading of the disk. This behavior implies that winds driven by external FUV radiation are more efficient in the disk dispersal in comparison to a similar disk subject to the radiation field of the host star. This trend, however, has already been  found by  AAC2013 in the case of the single star disks. We now find that  it also persists in the case of circumbinary disks. On the bottom row, the evolution of the disk mass (left) and the corresponding tracks in the plane of disk mass and radius (right) are shown. Disk dispersal in the case with external FUV radiation driven winds happens on a timescale by a factor of about 10 shorter than a similar disk with winds driven by host star X-ray radiation. We also note that \cite{Alexander12} who studied circumbinary disk evolution with winds driven by the host star radiation found a reduction in the disk lifetime. We, however, think that the reduction factor is overestimated due to the inconsistency between the implemented initial surface density and the reported initial disk mass. Furthermore, \cite{Alexander12} considered binaries with the larger separations that may affect disk lifetime.

\section{Astrophysical Implications}
In this section, we provide a comparison between the observed circumbinary disk mass and radius and our theoretical tracks in the mass and radius plane. In Table \ref{tab:2}, characteristics of a few binary systems and their associated circumbinary disks where to reside in different star forming regions are presented. In cluster with a sufficient number of stars, there are several O or B stars which can radiate FUV photons. For example, AK Sco resides in the upper Centaurus–Lupus star forming region which is a subgroup of the Scorpius–Centaurus association \citep{Andersen1989}. This association is the nearest OB association to us and contains several OB stars \citep{Blaauw1946, Blaauw1964}. Therefore, we can expect that circumbinary disks residing in this region are affected by the external FUV radiation driven winds.

All the model parameters cannot be constrained by the observational data, however, we can assume that the total binary mass is 1 or 2 solar mass because the mass of the observed binaries lies in this range. We neglect induced photoevaporative winds by the host stars and the external wind parameters are adopted as before. The initial disk total mass, as before, is assumed to be $M_{d0}=0.1 M_{\rm c}$.

\begin{table*}
\begin{center}
\caption{Properties of the binary systems and their disks. }\label{tab:2}
\begin{tabular}{||c|c|c|c|c|c|c|c|c||}\hline
Binary System & $r_{\rm d}[\rm{AU}]$ & $M_{\rm d}[\rm M_{\odot}]$ & $M_{\rm p}[\rm M_{\odot}]$ & $M_{\rm s}[\rm M_{\odot}]$ & $M_{\rm c}[\rm M_{\odot}]$ & $q(=M_{\rm s}/M_{\rm p})$ & References\\ \hline
AK Sco & 40 & 0.005 & 1.35 & 1.33 & 2.68 & 0.98 &  (1)\\
DQ Tau & 50 & 0.002-0.02 & 0.63 & 0.59 & 1.22 & 0.93 & (2), (3)\\
FS Tau A & 630 & 0.002 & 0.6 & 0.28 & 0.88 & 0.46 & (4), (5)\\
GG Tau A & 800 & 0.128 & 0.78 & 0.68 & 1.46 & 0.87 & (6), (7)\\
GG Tau Ab & 13 & - & 0.38 & 0.3 & 0.68 & 0.79 & (8), (9)\\
HH 30 & 250 & 0.004 & 0.31 & 0.14 & 0.45 & 0.45 & (10), (11)\\
L1165-SMM1 & 100 & 0.03 & - & - & 0.1-0.25 & - & (12)\\
L1551 NE & 300 & 0.043 & - & - & 0.8 & 0.19 & (13), (14), (15)\\
UY Aur & 2100 & 1.2 & - & - & 1.73 & - & (16), (17)\\
V4046 Sgr & 350 & 0.09 & 0.9 & 0.85 & 1.75 & 0.94 & (18), (19), (20)\\
UZ Tau E & - & 0.063 & 1.0 & 0.3 & 1.3 & 0.3 & (21), (22)\\ \hline
\end{tabular}
\end{center}
\begin{tablenotes}
   \small
   \item \textbf{Notes}: Column 1: Source. Columns 2: Disk radius. Column 3: Disk mass. Column 4: Mass of primary star. Column 5: Mass of secondary star. Column 6: Total mass of host stars. Column 7: Mass ratio. \\
      \textbf{References}: (1)\cite{Alencar2003}; (2)\cite{Mathieu1997}; (3)\cite{ Czekala2016}; (4)\cite{Andrews2005}; (5)\cite{Hioki2011}; (6)\cite{Guilloteau1999}; (7)\cite{White1999}; (8)\cite{diFolco2014}; (9)\cite{Yang2017}; (10)\cite{Pety2006}; (11)\cite{Estalella2012}; (12)\cite{Tobin2013}; (13)\cite{Takakuwa2012}; (14)\cite{Takakuwa2015}; (15)\cite{Lim2016}; (16)\cite{Duvert1998}; (17)\cite{Hioki2007}; (18)\cite{Rodriguez2010}; (19)\cite{Rosenfeld2012}; (20)\cite{Rosenfeld2013}; (21)\cite{Prato2002}; (22)\cite{Jensen2007}.
\end{tablenotes}

\end{table*}

\begin{figure}
\includegraphics[scale=0.5]{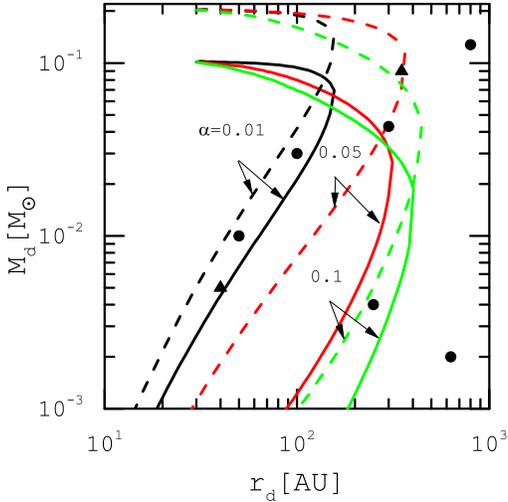}
\caption{Theoretical tracks in the disk mass and radius plane for different values of the viscosity parameter, as labeled. The wind parameters are as before and rest of the input parameters are $q=f=1.0$, $a_{\rm b}=0.2$ AU and $s=1/2$. Filled circles and triangles represent the observational data of Table \ref{tab:2} and correspond to the disks with $M_{\rm c}$ close to 1 and 2 ${\rm M}_{\odot}$. The solid and dashed curves correspond to the binaries with the total mass $M_{\rm c}=1.0 $ ${\rm M}_{\odot}$ and $M_{\rm c}=2.0$ ${\rm M}_{\odot}$, respectively.} \label{fig:f11}
\end{figure}

\begin{figure*}
\includegraphics[scale=1.0]{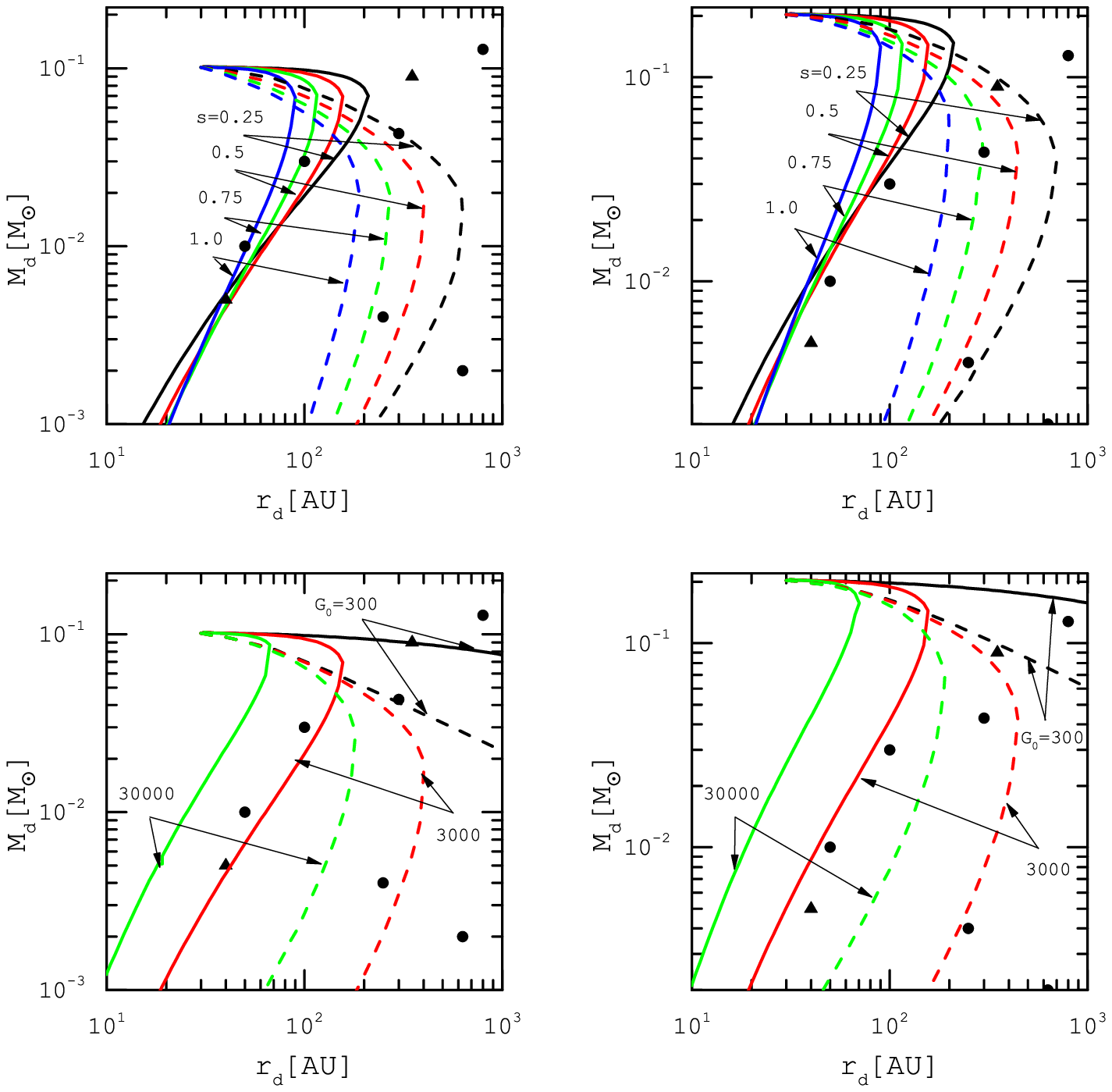}
\caption{Evolutionary tracks in the mass and radius plane for circumbinary disks with $\alpha=0.01$ (solid curve) and $\alpha=0.1$ (dashed curve) along with the observational data of Table \ref{tab:2}. The left-hand panels display evolutionary tracks for $M_{\rm c}=1.0~{\rm M}_{\odot}$, whereas in the right-hand panels the binary total mass is $M_{\rm c}=2.0~{\rm M}_{\odot}$. On the top panels, the evolutionary tracks of the circumbinary disks are shown for different values of $s$, as labeled. Role of the parameter $G_0$ in the evolutionary tracks is shown in the bottom panels. }\label{fig:f12}
\end{figure*}

\begin{figure*}
\includegraphics[scale=1.0]{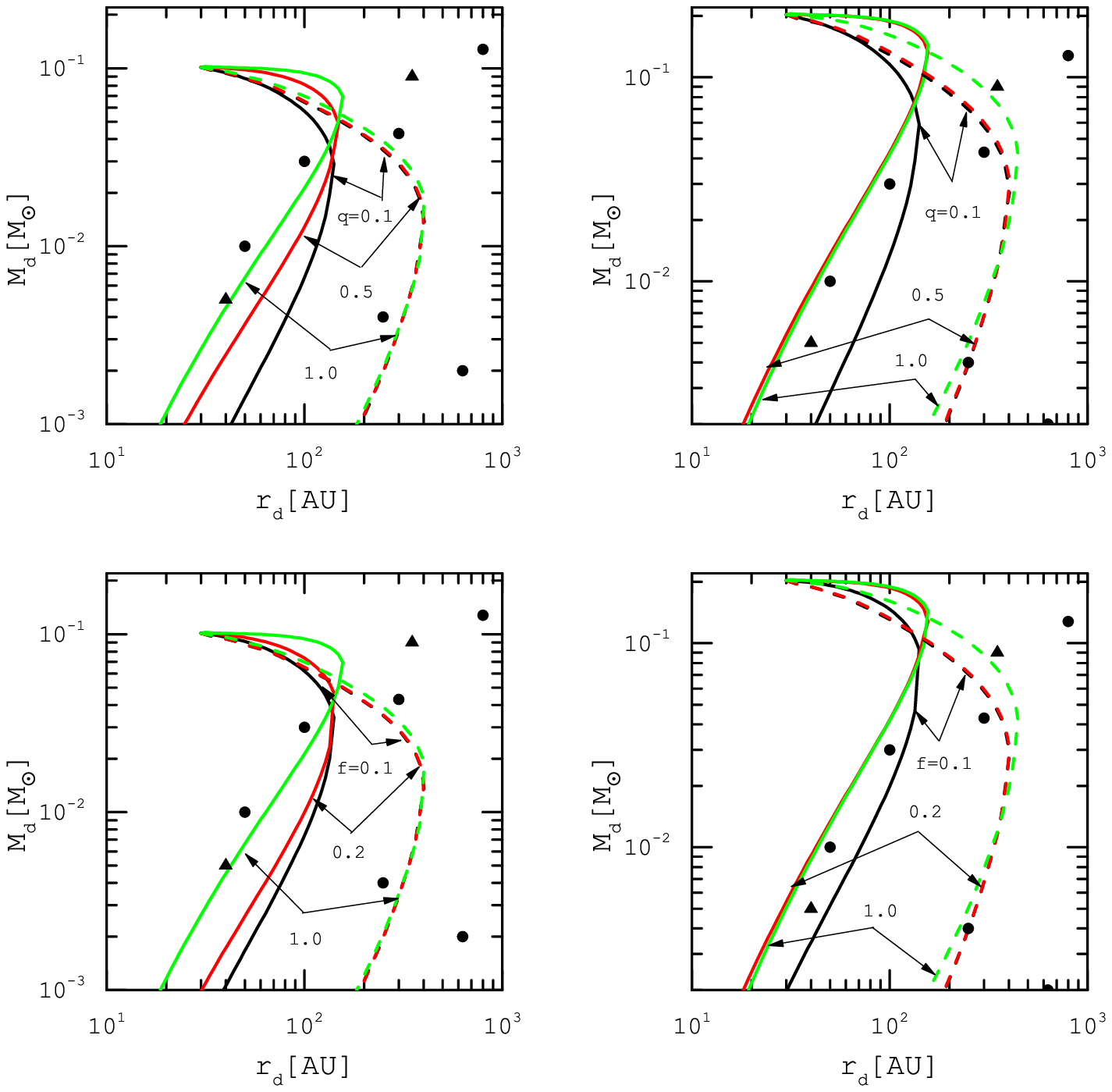}
 \caption{Similar to Figure \ref{fig:f12}, but for $s=1/2$ and $G_0=3000$ and different values of binary parameters. Role of the parameters $q$ and $f$ in the evolutionary tracks is illustrated in the top and  bottom panels, respectively. }\label{fig:f13}
\end{figure*}

Figures \ref{fig:f11}, \ref{fig:f12}, and \ref{fig:f13} show theoretical tracks in the disk mass and radius for $M_{\rm c}=1 $ ${\rm M}_{\odot}$ and  $M_{\rm c}=2 $ ${\rm M}_{\odot}$ with the observational data shown as filled circles and triangles. Note that filled circles and triangles correspond to the circumbinary disks with $M_{\rm c}$ close to 1 and 2 ${\rm M}_{\odot}$, respectively. In Figure \ref{fig:f11}, we have $q=f=1.0$, $a_{\rm b}=0.2$ AU, $G_0=3000$, and $s=1/2$ and the corresponding evolutionary tracks in the mass and radius plane are shown for the different viscosity coefficients, as labeled. Figure \ref{fig:f11} indicates that the viscosity coefficient is between 0.01 and 0.1 to have evolutionary tracks consistent with the observed circumbinary disks listed in Table \ref{tab:2}. Disk lifetime, therefore, is estimated to be in the ranges 0.14-1.2 Myr and 0.25-1.7 Myr for the total binary mass 1 ${\rm M}_{\odot}$ and 2 ${\rm M}_{\odot}$, respectively.

Figure \ref{fig:f12} displays the role of parameters $s$ (top panels) and $G_0$ (bottom panels) on the evolutionary tracks of equal-mass binaries with $f=1.0$. On the top panels, the evolutionary tracks are shown for different values of the temperature exponent $s$, as labeled. For $\alpha=0.1$ and the temperature exponent within the range, $s=0.25$ and $0.5$, the evolutionary tracks are nearly consistent with the observed values. For a lower viscosity coefficient $\alpha=0.01$, however, a reasonable agreement is achieved if the temperature exponent lies in a range between 0.75 and 1. The bottom panels illustrate the effect of flux level $G_0$ on the evolutionary tracks. For the circumbinary disks exposed to the radiation field with $G_0=3000$, we find that these tracks are closer to the observed binaries.

In Figure \ref{fig:f13}, we investigate how the rest of model parameters including the mass ratio $q$ and the binary torque coefficient $f$ can affect theoretical tracks in the disk mass and radius plane in comparison to the available observational data. In the left-hand panels, the binary total mass is 1 ${\rm M}_{\odot}$, whereas the right-hand panels correspond to the total mass 2 ${\rm M}_{\odot}$. 
 
The top panels display a comparison between observational data and our evolutionary tracks for different values of the mass ratio $q$. The temperature exponent and the binary torque coefficient are $s=1/2$ and $f=1.0$, respectively. The best agreement is achieved for a model with $q=1.0$. Most of the binaries in Table \ref{tab:2} are nearly equal-mass binaries with a  mass ratio close to unity. Therefore, we find that their lifetimes are within $(1-2) \times 10^6$ yr for viscosity coefficient $\alpha =0.01$, and their lifetimes are   extended to $(1.4-2.5) \times 10^6$ yr if a lower viscosity $\alpha=0.01$ is adopted. It is worth noting that our estimates of the lifetime depend on the binary total mass and the initial disk mass. Finally, in the bottom plots of Figure \ref{fig:f13} the evolutionary tracks for different values of the binary torque coefficient $f$ are displayed. Here, we have $s=1/2$ and $q=1.0$. If we set $M_{\rm c}=1.0$ M$_{\odot}$, our model for $f=1.0$ is in good agreement with the observed disks, whereas for $M_{\rm c}=2.0$ M$_{\odot}$, the agreement is achieved for the binary torque coefficient in an interval from 0.1 to 1.0. Therefore, our calculations suggest that the circumbinary disks with $M_{\rm c}=2.0$ M$_{\odot}$ can survive during $(0.9-1.7)\times 10^6$ yr.

\section{discussion}
Photoevaporative winds driven by radiation of the host stars or ambient sources are very effective in the disk mass removal. We studied the evolution of a circumbinary PPD with the induced wind due to  external FUV radiation field. Therefore, an obvious feature is the emergence of a shrinking outer radius, which leads to the disk dispersal via outside-in  clearing. A circumbinary PPD evolution is found to be significantly dominated by the binary torque, however, its viscous spreading is suppressed by the photoevaporative winds which control the disk clearing. Although the  significance of the binary torque is restricted to the innermost region, the resulting mass pileup in this region and its subsequent viscous spreading would influence disk structure entirely.

We implemented simplifying assumptions to focus on the main aspects of  the circumbinary PPDs evolution in the presence of winds. For example, energy balance through the disk has not been included. We thus used  a power-law function of the radial distance for the disk temperature distribution. The exponent of this profile is a model parameter that controls the steepness of the disk temperature. We found that lifetimes of the circumbinary PPDs increase as the disk temperature becomes steeper. However, a single temperature exponent is not adequate to describe temperature profile through the disk because there are different heating sources that we did not consider. While viscous heating is a dominant heating mechanism in the innermost regions, stellar irradiation becomes a dominant heating source in the outer disk parts. Although implementing a more realistic model with the energy equation and the associated heating and cooling mechanisms are needed for obtaining disk temperature self-consistently, our main conclusions do not change because we explored disk properties for a broad range of the temperature exponent. As we mentioned earlier, the binary torque strongly suppress the mass accretion rate at the disk inner edge which leads to the mass pileup.Some amount of accretion may still proceed at the inner edge, however, its rate strongly depends on the binary torque and disk turbulence that are not well constrained. Since this effect has already been implemented by \cite{Alexander12}, we did not incorporate it in our analysis and our focus was instead an extreme situation with negligible mass accretion rate at the inner edge. Nevertheless, we did some disk evolution calculations and  found that, for instance, when the accretion rate near to the inner edge is allowed to be enhanced by about 10 percent, disk lifetime reduces around 10 percent. This effect does not change our main results significantly.  

Although the disk turbulence is commonly believed to be driven by MRI, detailed models that include ionization sources show that MRI can not operate in the entire disk and a "dead zone" is created at intermediate radial distances where is shielded from ionizing radiation \citep{Gammie96}. In PPDs which contain dead-zones, therefore, accretion  proceeds in a layered fashion which means that the flow of matter occurs in the outer layers surrounding the dead zones whereas in other disk parts magnetic coupling is strong enough to trigger MRI as the main agent of the accretion \citep[e.g.,][]{Turner2008,martin12,Dzy}. The strength of turbulence in the dead zone is thought to be weaker in comparison to the disk regions where magnetically are active. In PPDs with dead-zones, we expect the viscosity coefficient to vary with the radial distance \citep[e.g.,][]{Mohanty2018}. Various features of PPDs including motion of dust particles and planetesimals \citep[e.g.,][]{Okuzumi2013,Chatterjee2014}, planet formation and its migration are dramatically affected in the presence of the dead zone \citep[e.g.,][]{Faure2016}. Therefore, determining location and size of the dead zone in the PPDs is an important problem.   
Since the dead zone with low viscosity can slow down the accretion rate, the pile-up of dust and gas occurs at the inner dead zone boundary between magnetically active and non-actives regions \citep{Chatterjee2014}. 

Most prior works, however, were focused on the dead zone properties and its physical consequences in single star disks \citep[e.g.,][]{fromang2002,Matsumura,Turner2008,martin12,Dzy}. We showed that a photoevaporative circumbinary disk evolves with a higher surface density at a given time and radius in comparison to its circumstellar disk counterpart. Therefore, ionization radiation is expected to penetrate a shorter distance into a circumbinary disk comparing to a similar single star disk. In other words, the penetration power of the ionization radiation is weaker in the circumbinary disks due to their enhanced surface density. We did not calculate location and size of the dead zone in the photoevaporative circumbinary disks, however, our findings provide this qualitative insight that dead zones in these systems are bigger in comparison to the similar circumstellar disks. The dead zone in a PPD is expected to gradually shrink with the disk dispersal because of the wind mass-loss. In the circumbinary disks, however, this process can happen over a longer time  because the circumbinary lifetime is longer than a similar circumstellar disk. Therefore, dead zones in the photoevaporative circumbinary disks are larger and can survive over a longer time in comparison to the similar single star disks. 

We, however, did not quantify dead zone size and its lifetime in the circumbinary PPDs with photoevaporative winds and this problem needs to be studied further. A bigger dead zone with a longer lifetime is expected to dramatically affect not only grains motion through the dead zone and their pileup at its inner edge but also other related features including planet formation in this region. 

Photoevaporative winds are generally efficient in mass removal from the dust-free disk surface layers. However, a dusty layer can be formed at the disk midplane due to quick dust settling. Therefore, the disk dust-to-gas density  increases in the presence of photoevaporative winds \citep{Carrera2017,Ercolano2017}.

Instabilities in the PPDs due to the dust and gas interaction such as the streaming instability \citep[SI;][]{Youdin2005} and the secular gravitational instability \citep[SGI;][]{Youdin2011,shariff,Takahashi2014} are  believed to seed planetesimal formation. Growth rates of these instabilities strongly depend on the dust-to-gas density ratio. This ratio in the PPDs is estimated to be  about 0.01 which is much smaller than that  required  to trigger SI or SGI efficiently. For instance, the fastest growth rate due to SI happens when the dust-to-gas density ratio is of order unity \citep[][]{Youdin2011}. Any mechanism, including mass-loss by winds, that may enhance the dust-to-gas density ratio is needed to promote the instabilities. We showed that a circumbinary disk evolves with a larger surface density in comparison to the circumstellar disk counterpart. Since dust abundance is not affected by the wind mass-loss, the dust-to-gas density ratio  in the photoevaporative circumstellar disks is expected to increase more than  similar circumbinary disks. We did not quantify this theoretical expectation because dust dynamics has not been included in our model. We, therefore, suggest it as a direction for future study. Recent studies have calculated dust-to-gas density ratio in the circumstellar disks with the photoevaporative winds \citep{Carrera2017,Ercolano2017}. On the other hand, disk lifetime is a severe constraint and provides an upper limit to the growth timescale of the instabilities. Although enhancement of the dust-to-gas density in the circumbinary disks is expected to be smaller than that in the similar circumstellar disk that leads to a slower growth rate of the instabilities, the circumbinary disks can survive on a long time and the instabilities have more time to grow efficiently. These theoretical expectations need to be explored further. 

Our work helps clarify how winds induced by external FUV radiation field can influence the structure of the circumbinary PPDs. From this study, we conclude:

- Mass removal by external FUV radiation field is more efficient than winds induces by the host star radiation. We thus find that circumbinary PPDs where to reside in stars associations are generally dispersed during a shorter time in comparison to the similar circumbinary PPDs where are in isolation. This is a conclusion in agreement with work by AAC2013 in the case of single star disks.

- In the presence of photoevaporative winds, the circumbinary PPDs are dispersed over a time by a factor of two or more longer than  the circumstellar disks analogous.   

- This lifetime enhancement happens due to mass pileup by the binary torque at the disk innermost regions. 

- Lifetime dependence on the viscosity coefficient in the circumbinary PPDs is stronger than similar single star disks. 

- Disk lifetime increases as the radial temperature distribution becomes steeper.

\section*{Acknowledgements}

We are grateful to the referee for a very constructive and thoughtful report that greatly helped us to improve the paper. This work has been supported financially by the Research Institute for Astronomy \& Astrophysics of Maragha (RIAAM) under
research project No. 1/5750-11. 

\bibliographystyle{apj}
\bibliography{reference} 

\begin{thebibliography}{}

\bibitem[\protect\citeauthoryear{{Adams} et~al.}{{Adams}
  et~al.}{2004}]{Adams2004}
{Adams}, F.~C., {Hollenbach}, D., {Laughlin}, G.,  \& {Gorti}, U. 2004, \apj,
  611, 360

\bibitem[\protect\citeauthoryear{{Alencar} et~al.}{{Alencar}
  et~al.}{2003}]{Alencar2003}
{Alencar}, S.~H.~P., {Melo}, C.~H.~F., {Dullemond}, C.~P., {Andersen}, J.,
  {Batalha}, C., {Vaz}, L.~P.~R.,  \& {Mathieu}, R.~D. 2003, \aap, 409, 1037

\bibitem[\protect\citeauthoryear{{Alexander}}{{Alexander}}{2012}]{Alexander12}
{Alexander}, R. 2012, \apjl, 757, L29

\bibitem[\protect\citeauthoryear{{Alexander}, {Clarke}, \&
  {Pringle}}{{Alexander} et~al.}{2006}]{Alex2006}
{Alexander}, R.~D., {Clarke}, C.~J.,  \& {Pringle}, J.~E. 2006, \mnras, 369,
  229

\bibitem[\protect\citeauthoryear{{Andersen} et~al.}{{Andersen}
  et~al.}{1989}]{Andersen1989}
{Andersen}, J., {Lindgren}, H., {Hazen}, M.~L.,  \& {Mayor}, M. 1989, \aap,
  219, 142

\bibitem[\protect\citeauthoryear{{Anderson}, {Adams}, \& {Calvet}}{{Anderson}
  et~al.}{2013}]{Ander13}
{Anderson}, K.~R., {Adams}, F.~C.,  \& {Calvet}, N. 2013, \apj, 774, 9

\bibitem[\protect\citeauthoryear{{Andrews} \& {Williams}}{{Andrews} \&
  {Williams}}{2005}]{Andrews2005}
{Andrews}, S.~M.,  \& {Williams}, J.~P. 2005, \apj, 631, 1134

\bibitem[\protect\citeauthoryear{{Armitage}}{{Armitage}}{2011}]{Armitage11}
{Armitage}, P.~J. 2011, \araa, 49, 195

\bibitem[\protect\citeauthoryear{{Armitage} \& {Natarajan}}{{Armitage} \&
  {Natarajan}}{2002}]{Armitage2002}
{Armitage}, P.~J.,  \& {Natarajan}, P. 2002, \apjl, 567, L9

\bibitem[\protect\citeauthoryear{{Bai} et~al.}{{Bai} et~al.}{2016}]{Bai2016}
{Bai}, X.-N., {Ye}, J., {Goodman}, J.,  \& {Yuan}, F. 2016, \apj, 818, 152

\bibitem[\protect\citeauthoryear{{Balbus} \& {Hawley}}{{Balbus} \&
  {Hawley}}{1991}]{Balbus91}
{Balbus}, S.~A.,  \& {Hawley}, J.~F. 1991, \apj, 376, 214

\bibitem[\protect\citeauthoryear{{Blaauw}}{{Blaauw}}{1946}]{Blaauw1946}
{Blaauw}, A. 1946, Publications of the Kapteyn Astronomical Laboratory
  Groningen, 52, 1

\bibitem[\protect\citeauthoryear{{Blaauw}}{{Blaauw}}{1964}]{Blaauw1964}
{Blaauw}, A. 1964, \araa, 2, 213

\bibitem[\protect\citeauthoryear{{Blandford} \& {Payne}}{{Blandford} \&
  {Payne}}{1982}]{Bland82}
{Blandford}, R.~D.,  \& {Payne}, D.~G. 1982, \mnras, 199, 883

\bibitem[\protect\citeauthoryear{{Carrera} et~al.}{{Carrera}
  et~al.}{2017}]{Carrera2017}
{Carrera}, D., {Gorti}, U., {Johansen}, A.,  \& {Davies}, M.~B. 2017, \apj,
  839, 16

\bibitem[\protect\citeauthoryear{{Chatterjee} \& {Tan}}{{Chatterjee} \&
  {Tan}}{2014}]{Chatterjee2014}
{Chatterjee}, S.,  \& {Tan}, J.~C. 2014, \apj, 780, 53

\bibitem[\protect\citeauthoryear{{Clarke}}{{Clarke}}{2007}]{Clarke2007}
{Clarke}, C.~J. 2007, \mnras, 376, 1350

\bibitem[\protect\citeauthoryear{{Czekala} et~al.}{{Czekala}
  et~al.}{2016}]{Czekala2016}
{Czekala}, I., {Andrews}, S.~M., {Torres}, G., {Jensen}, E.~L.~N., {Stassun},
  K.~G., {Wilner}, D.~J.,  \& {Latham}, D.~W. 2016, \apj, 818, 156

\bibitem[\protect\citeauthoryear{{Di Folco} et~al.}{{Di Folco}
  et~al.}{2014}]{diFolco2014}
{Di Folco}, E., et~al. 2014, \aap, 565, L2

\bibitem[\protect\citeauthoryear{{Doyle} et~al.}{{Doyle}
  et~al.}{2011}]{Doyle11}
{Doyle}, L.~R., et~al. 2011, Science, 333, 1602

\bibitem[\protect\citeauthoryear{{Duvert} et~al.}{{Duvert}
  et~al.}{1998}]{Duvert1998}
{Duvert}, G., {Dutrey}, A., {Guilloteau}, S., {Menard}, F., {Schuster}, K.,
  {Prato}, L.,  \& {Simon}, M. 1998, \aap, 332, 867

\bibitem[\protect\citeauthoryear{{Dzyurkevich} et~al.}{{Dzyurkevich}
  et~al.}{2013}]{Dzy}
{Dzyurkevich}, N., {Turner}, N.~J., {Henning}, T.,  \& {Kley}, W. 2013, \apj,
  765, 114

\bibitem[\protect\citeauthoryear{{Ercolano} et~al.}{{Ercolano}
  et~al.}{2017}]{Ercolano2017}
{Ercolano}, B., {Jennings}, J., {Rosotti}, G.,  \& {Birnstiel}, T. 2017,
  \mnras, 472, 4117

\bibitem[\protect\citeauthoryear{{Ercolano} \& {Pascucci}}{{Ercolano} \&
  {Pascucci}}{2017}]{Ercolano17}
{Ercolano}, B.,  \& {Pascucci}, I. 2017, Royal Society Open Science, 4, 170114

\bibitem[\protect\citeauthoryear{{Estalella} et~al.}{{Estalella}
  et~al.}{2012}]{Estalella2012}
{Estalella}, R., {L{\'o}pez}, R., {Anglada}, G., {G{\'o}mez}, G., {Riera}, A.,
  \& {Carrasco-Gonz{\'a}lez}, C. 2012, \aj, 144, 61

\bibitem[\protect\citeauthoryear{{Faure} \& {Nelson}}{{Faure} \&
  {Nelson}}{2016}]{Faure2016}
{Faure}, J.,  \& {Nelson}, R.~P. 2016, \aap, 586, A105

\bibitem[\protect\citeauthoryear{{Frank}, {King}, \& {Raine}}{{Frank}
  et~al.}{2002}]{Frank2002}
{Frank}, J., {King}, A.,  \& {Raine}, D.~J. 2002, {Accretion Power in
  Astrophysics: Third Edition} 398

\bibitem[\protect\citeauthoryear{{Fromang}, {Terquem}, \& {Balbus}}{{Fromang}
  et~al.}{2002}]{fromang2002}
{Fromang}, S., {Terquem}, C.,  \& {Balbus}, S.~A. 2002, \mnras, 329, 18

\bibitem[\protect\citeauthoryear{{Gammie}}{{Gammie}}{1996}]{Gammie96}
{Gammie}, C.~F. 1996, \apj, 457, 355

\bibitem[\protect\citeauthoryear{{Gorti}, {Dullemond}, \& {Hollenbach}}{{Gorti}
  et~al.}{2009}]{Gorti2009}
{Gorti}, U., {Dullemond}, C.~P.,  \& {Hollenbach}, D. 2009, \apj, 705, 1237

\bibitem[\protect\citeauthoryear{{Guilloteau}, {Dutrey}, \&
  {Simon}}{{Guilloteau} et~al.}{1999}]{Guilloteau1999}
{Guilloteau}, S., {Dutrey}, A.,  \& {Simon}, M. 1999, \aap, 348, 570

\bibitem[\protect\citeauthoryear{{Hioki} et~al.}{{Hioki}
  et~al.}{2011}]{Hioki2011}
{Hioki}, T., {Itoh}, Y., {Oasa}, Y., {Fukagawa}, M.,  \& {Hayashi}, M. 2011,
  \pasj, 63, 543

\bibitem[\protect\citeauthoryear{{Hioki} et~al.}{{Hioki}
  et~al.}{2007}]{Hioki2007}
{Hioki}, T., et~al. 2007, \aj, 134, 880

\bibitem[\protect\citeauthoryear{{Hollenbach} et~al.}{{Hollenbach}
  et~al.}{1994}]{Hollenbach1994}
{Hollenbach}, D., {Johnstone}, D., {Lizano}, S.,  \& {Shu}, F. 1994, \apj, 428,
  654

\bibitem[\protect\citeauthoryear{{Jensen} et~al.}{{Jensen}
  et~al.}{2007}]{Jensen2007}
{Jensen}, E.~L.~N., {Dhital}, S., {Stassun}, K.~G., {Patience}, J., {Herbst},
  W., {Walter}, F.~M., {Simon}, M.,  \& {Basri}, G. 2007, \aj, 134, 241

\bibitem[\protect\citeauthoryear{{Kimura}, {Kunitomo}, \& {Takahashi}}{{Kimura}
  et~al.}{2016}]{Kimura2016}
{Kimura}, S.~S., {Kunitomo}, M.,  \& {Takahashi}, S.~Z. 2016, \mnras, 461, 2257

\bibitem[\protect\citeauthoryear{{Kocsis}, {Haiman}, \& {Loeb}}{{Kocsis}
  et~al.}{2012}]{Kocsis2012}
{Kocsis}, B., {Haiman}, Z.,  \& {Loeb}, A. 2012, \mnras, 427, 2660

\bibitem[\protect\citeauthoryear{{Kratter} \& {Lodato}}{{Kratter} \&
  {Lodato}}{2016}]{Kratter2016}
{Kratter}, K.,  \& {Lodato}, G. 2016, \araa, 54, 271

\bibitem[\protect\citeauthoryear{{Kraus} et~al.}{{Kraus}
  et~al.}{2012}]{Kraus2012}
{Kraus}, A.~L., {Ireland}, M.~J., {Hillenbrand}, L.~A.,  \& {Martinache}, F.
  2012, \apj, 745, 19

\bibitem[\protect\citeauthoryear{{Li} \& {Sui}}{{Li} \& {Sui}}{2017}]{Li2017}
{Li}, M.,  \& {Sui}, N. 2017, \mnras, 466, 1205

\bibitem[\protect\citeauthoryear{{Li} \& {Xiao}}{{Li} \& {Xiao}}{2016}]{Li2016}
{Li}, M.,  \& {Xiao}, L. 2016, \apj, 820, 36

\bibitem[\protect\citeauthoryear{{Lim} et~al.}{{Lim} et~al.}{2016}]{Lim2016}
{Lim}, J., {Hanawa}, T., {Yeung}, P.~K.~H., {Takakuwa}, S., {Matsumoto}, T.,
  \& {Saigo}, K. 2016, \apj, 831, 90

\bibitem[\protect\citeauthoryear{{Liu} \& {Shapiro}}{{Liu} \&
  {Shapiro}}{2010}]{Shapiro2010}
{Liu}, Y.~T.,  \& {Shapiro}, S.~L. 2010, \prd, 82, 123011

\bibitem[\protect\citeauthoryear{{Lynden-Bell} \& {Pringle}}{{Lynden-Bell} \&
  {Pringle}}{1974}]{Lyn74}
{Lynden-Bell}, D.,  \& {Pringle}, J.~E. 1974, \mnras, 168, 603

\bibitem[\protect\citeauthoryear{{MacFadyen} \&
  {Milosavljevi{\'c}}}{{MacFadyen} \&
  {Milosavljevi{\'c}}}{2008}]{MacFadyen2008}
{MacFadyen}, A.~I.,  \& {Milosavljevi{\'c}}, M. 2008, \apj, 672, 83

\bibitem[\protect\citeauthoryear{{Martin}, {Armitage}, \& {Alexander}}{{Martin}
  et~al.}{2013}]{Martin13}
{Martin}, R.~G., {Armitage}, P.~J.,  \& {Alexander}, R.~D. 2013, \apj, 773, 74

\bibitem[\protect\citeauthoryear{{Martin} et~al.}{{Martin}
  et~al.}{2012}]{martin12}
{Martin}, R.~G., {Lubow}, S.~H., {Livio}, M.,  \& {Pringle}, J.~E. 2012,
  \mnras, 420, 3139

\bibitem[\protect\citeauthoryear{{Mathieu} et~al.}{{Mathieu}
  et~al.}{1997}]{Mathieu1997}
{Mathieu}, R.~D., {Stassun}, K., {Basri}, G., {Jensen}, E.~L.~N.,
  {Johns-Krull}, C.~M., {Valenti}, J.~A.,  \& {Hartmann}, L.~W. 1997, \aj, 113,
  1841

\bibitem[\protect\citeauthoryear{{Matsumura}, {Pudritz}, \&
  {Thommes}}{{Matsumura} et~al.}{2009}]{Matsumura}
{Matsumura}, S., {Pudritz}, R.~E.,  \& {Thommes}, E.~W. 2009, \apj, 691, 1764

\bibitem[\protect\citeauthoryear{{Miranda}, {Mu{\~n}oz}, \& {Lai}}{{Miranda}
  et~al.}{2017}]{Miranda2017}
{Miranda}, R., {Mu{\~n}oz}, D.~J.,  \& {Lai}, D. 2017, \mnras, 466, 1170

\bibitem[\protect\citeauthoryear{{Mohanty} et~al.}{{Mohanty}
  et~al.}{2018}]{Mohanty2018}
{Mohanty}, S., {Jankovic}, M.~R., {Tan}, J.~C.,  \& {Owen}, J.~E. 2018, \apj,
  861, 144

\bibitem[\protect\citeauthoryear{{Okuzumi} \& {Ormel}}{{Okuzumi} \&
  {Ormel}}{2013}]{Okuzumi2013}
{Okuzumi}, S.,  \& {Ormel}, C.~W. 2013, \apj, 771, 43

\bibitem[\protect\citeauthoryear{{Orosz} et~al.}{{Orosz}
  et~al.}{2012}]{Orosz12}
{Orosz}, J.~A., et~al. 2012, Science, 337, 1511

\bibitem[\protect\citeauthoryear{{Owen}, {Ercolano}, \& {Clarke}}{{Owen}
  et~al.}{2011a}]{Owen11}
{Owen}, J.~E., {Ercolano}, B.,  \& {Clarke}, C.~J. 2011a, \mnras, 412, 13

\bibitem[\protect\citeauthoryear{{Owen}, {Ercolano}, \& {Clarke}}{{Owen}
  et~al.}{2011b}]{Owen2011}
{Owen}, J.~E., {Ercolano}, B.,  \& {Clarke}, C.~J. 2011b, \mnras, 412, 13

\bibitem[\protect\citeauthoryear{{Owen} et~al.}{{Owen} et~al.}{2010a}]{Owen10}
{Owen}, J.~E., {Ercolano}, B., {Clarke}, C.~J.,  \& {Alexander}, R.~D. 2010a,
  \mnras, 401, 1415

\bibitem[\protect\citeauthoryear{{Owen} et~al.}{{Owen}
  et~al.}{2010b}]{Owen2010}
{Owen}, J.~E., {Ercolano}, B., {Clarke}, C.~J.,  \& {Alexander}, R.~D. 2010b,
  \mnras, 401, 1415

\bibitem[\protect\citeauthoryear{{Pelupessy} \& {Portegies Zwart}}{{Pelupessy}
  \& {Portegies Zwart}}{2013}]{Pelupessy2013}
{Pelupessy}, F.~I.,  \& {Portegies Zwart}, S. 2013, \mnras, 429, 895

\bibitem[\protect\citeauthoryear{{Pety} et~al.}{{Pety} et~al.}{2006}]{Pety2006}
{Pety}, J., {Gueth}, F., {Guilloteau}, S.,  \& {Dutrey}, A. 2006, \aap, 458,
  841

\bibitem[\protect\citeauthoryear{{Prato} et~al.}{{Prato}
  et~al.}{2002}]{Prato2002}
{Prato}, L., {Simon}, M., {Mazeh}, T., {Zucker}, S.,  \& {McLean}, I.~S. 2002,
  \apjl, 579, L99

\bibitem[\protect\citeauthoryear{{Rafikov}}{{Rafikov}}{2013}]{Rafikov2013}
{Rafikov}, R.~R. 2013, \apj, 774, 144

\bibitem[\protect\citeauthoryear{{Ragusa}, {Lodato}, \& {Price}}{{Ragusa}
  et~al.}{2016}]{Ragusa2016}
{Ragusa}, E., {Lodato}, G.,  \& {Price}, D.~J. 2016, \mnras, 460, 1243

\bibitem[\protect\citeauthoryear{{Rodriguez} et~al.}{{Rodriguez}
  et~al.}{2010}]{Rodriguez2010}
{Rodriguez}, D.~R., {Kastner}, J.~H., {Wilner}, D.,  \& {Qi}, C. 2010, \apj,
  720, 1684

\bibitem[\protect\citeauthoryear{{Roedig} et~al.}{{Roedig}
  et~al.}{2012}]{Roedig2012}
{Roedig}, C., {Sesana}, A., {Dotti}, M., {Cuadra}, J., {Amaro-Seoane}, P.,  \&
  {Haardt}, F. 2012, \aap, 545, A127

\bibitem[\protect\citeauthoryear{{Rosenfeld} et~al.}{{Rosenfeld}
  et~al.}{2013}]{Rosenfeld2013}
{Rosenfeld}, K.~A., {Andrews}, S.~M., {Wilner}, D.~J., {Kastner}, J.~H.,  \&
  {McClure}, M.~K. 2013, \apj, 775, 136

\bibitem[\protect\citeauthoryear{{Rosenfeld} et~al.}{{Rosenfeld}
  et~al.}{2012}]{Rosenfeld2012}
{Rosenfeld}, K.~A., {Andrews}, S.~M., {Wilner}, D.~J.,  \& {Stempels}, H.~C.
  2012, \apj, 759, 119

\bibitem[\protect\citeauthoryear{{Rosotti} \& {Clarke}}{{Rosotti} \&
  {Clarke}}{2018}]{Rosotti18}
{Rosotti}, G.~P.,  \& {Clarke}, C.~J. 2018, \mnras, 473, 5630

\bibitem[\protect\citeauthoryear{{Schwamb} et~al.}{{Schwamb}
  et~al.}{2013}]{Schwamb13}
{Schwamb}, M.~E., et~al. 2013, \apj, 768, 127

\bibitem[\protect\citeauthoryear{{Shadmehri} \& {Khajenabi}}{{Shadmehri} \&
  {Khajenabi}}{2015}]{shadmehri2015}
{Shadmehri}, M.,  \& {Khajenabi}, F. 2015, \mnras, 447, 1439

\bibitem[\protect\citeauthoryear{{Shakura} \& {Sunyaev}}{{Shakura} \&
  {Sunyaev}}{1973}]{SS73}
{Shakura}, N.~I.,  \& {Sunyaev}, R.~A. 1973, \aap, 24, 337

\bibitem[\protect\citeauthoryear{{Shariff} \& {Cuzzi}}{{Shariff} \&
  {Cuzzi}}{2011}]{shariff}
{Shariff}, K.,  \& {Cuzzi}, J.~N. 2011, \apj, 738, 73

\bibitem[\protect\citeauthoryear{{Takahashi} \& {Inutsuka}}{{Takahashi} \&
  {Inutsuka}}{2014}]{Takahashi2014}
{Takahashi}, S.~Z.,  \& {Inutsuka}, S.-i. 2014, \apj, 794, 55

\bibitem[\protect\citeauthoryear{{Takakuwa} et~al.}{{Takakuwa}
  et~al.}{2015}]{Takakuwa2015}
{Takakuwa}, S., {Kiyokane}, K., {Saigo}, K.,  \& {Saito}, M. 2015, \apj, 814,
  160

\bibitem[\protect\citeauthoryear{{Takakuwa} et~al.}{{Takakuwa}
  et~al.}{2012}]{Takakuwa2012}
{Takakuwa}, S., {Saito}, M., {Lim}, J., {Saigo}, K., {Sridharan}, T.~K.,  \&
  {Patel}, N.~A. 2012, \apj, 754, 52

\bibitem[\protect\citeauthoryear{{Tang}, {MacFadyen}, \& {Haiman}}{{Tang}
  et~al.}{2017}]{Tang2017}
{Tang}, Y., {MacFadyen}, A.,  \& {Haiman}, Z. 2017, \mnras, 469, 4258

\bibitem[\protect\citeauthoryear{{Tobin} et~al.}{{Tobin}
  et~al.}{2013}]{Tobin2013}
{Tobin}, J.~J., et~al. 2013, \apj, 779, 93

\bibitem[\protect\citeauthoryear{{Turner} \& {Sano}}{{Turner} \&
  {Sano}}{2008}]{Turner2008}
{Turner}, N.~J.,  \& {Sano}, T. 2008, \apjl, 679, L131

\bibitem[\protect\citeauthoryear{{Vartanyan}, {Garmilla}, \&
  {Rafikov}}{{Vartanyan} et~al.}{2016}]{Vartan16}
{Vartanyan}, D., {Garmilla}, J.~A.,  \& {Rafikov}, R.~R. 2016, \apj, 816, 94

\bibitem[\protect\citeauthoryear{{Wang} \& {Goodman}}{{Wang} \&
  {Goodman}}{2017}]{wang2017ApJ}
{Wang}, L.,  \& {Goodman}, J.~J. 2017, \apj, 835, 59

\bibitem[\protect\citeauthoryear{{White} et~al.}{{White}
  et~al.}{1999}]{White1999}
{White}, R.~J., {Ghez}, A.~M., {Reid}, I.~N.,  \& {Schultz}, G. 1999, \apj,
  520, 811

\bibitem[\protect\citeauthoryear{{Xiao} \& {Chang}}{{Xiao} \&
  {Chang}}{2018}]{Xiao18}
{Xiao}, L.,  \& {Chang}, Q. 2018, \apj, 853, 22

\bibitem[\protect\citeauthoryear{{Yang} et~al.}{{Yang} et~al.}{2017}]{Yang2017}
{Yang}, Y., et~al. 2017, \aj, 153, 7

\bibitem[\protect\citeauthoryear{{Youdin}}{{Youdin}}{2011}]{Youdin2011}
{Youdin}, A.~N. 2011, \apj, 731, 99

\bibitem[\protect\citeauthoryear{{Youdin} \& {Goodman}}{{Youdin} \&
  {Goodman}}{2005}]{Youdin2005}
{Youdin}, A.~N.,  \& {Goodman}, J. 2005, \apj, 620, 459

\bibitem[\protect\citeauthoryear{{Yu}}{{Yu}}{2002}]{Yu2002}
{Yu}, Q. 2002, \mnras, 331, 935

\end{thebibliography}
\clearpage

\end{document}